\documentclass[12pt]{article}
\usepackage{a4wide}
\usepackage{psfrag}
\usepackage{graphics}
\usepackage{amsfonts}
\usepackage{amstext}
\usepackage{amsmath}
\usepackage{amssymb}
\usepackage{amsthm}
\usepackage{cases}
\usepackage[mathscr]{eucal}
\usepackage{cite}
\newtheorem{theorem}{Theorem}[section]
\newtheorem{proposition}[theorem]{Proposition}
\renewcommand{\theequation}{\arabic{section}.\arabic{equation}}
\newcommand{\pd}[1]{\frac{\partial}{\partial #1}}

\def\a{\alpha}

\def\Ai{{\rm Ai}}
\def\b{\beta}
\def\d{\delta}

\def\g{\gamma}
\def\G{\mathcal{G}}
\def\i{\infty}

\def\K{\mathcal{K}}
\def\l{\lambda}
\def\N{\mathbb{N}}
\def\o{\omega}
\def\p{\phi}

\def\P{\mathbb{P}}
\def\r{\rho}

\def\t{\tau}
\def\th{\theta}

\def\Z{\mathbb{Z}}

\begin{document}       
\title{\bf{Fluctuations of the one-dimensional polynuclear growth model 
with external sources}}

\author{
\renewcommand{\thefootnote}{\arabic{footnote}}
\vspace{5mm}
T. Imamura \footnotemark[1] ~and T. Sasamoto \footnotemark[2]
}


\author{
\vspace{5mm}
T. Imamura 
{\footnote {\tt e-mail: imamura@monet.phys.s.u-tokyo.ac.jp}}
~~and
T. Sasamoto 
{\footnote {\tt e-mail: sasamoto@stat.phys.titech.ac.jp}}
 \\
{\it$^*$Department of Physics, Graduate School of Science,}\\
{\it University of Tokyo,}\\
{\it Hongo 7-3-1, Bunkyo-ku, Tokyo 113-0033, Japan}\\
\vspace{0mm}\\
{\it$^{\dag}$Department of Physics, Tokyo Institute of Technology,}\\
{\it Oh-okayama 2-12-1, Meguro-ku, Tokyo 152-8551, Japan}\\
}

\date{} 
\maketitle

\begin{abstract}
The one-dimensional polynuclear growth model with external sources
at edges is studied. 
The height fluctuation at the origin is known to be given by 
either the Gaussian, the GUE Tracy-Widom distribution, 
or certain distributions called GOE$^2$ and $F_0$, 
depending on the strength of the sources.
We generalize these results and show that the scaling limit of 
the multi-point equal time height fluctuations of the model 
are described by the Fredholm determinant, of which the limiting 
kernel is explicitly obtained. 
In particular we obtain two new kernels, describing 
transitions between the above one-point distributions.
One expresses the transition from the GOE$^2$
to the GUE Tracy-Widom distribution or to the Gaussian;
the other the transition from $F_0$ to the Gaussian.
The results specialized to the fluctuation 
at the origin are shown to be equivalent to the previously 
obtained ones via the Riemann-Hilbert method.

\vspace{3mm}\noindent
[Keywords: polynuclear growth; KPZ universality class;  
random matrices; Tracy-Widom distribution; Airy process]


\end{abstract}
\newpage

\section{Introduction}
Surface growth is an interesting phenomenon in nature \cite{Me1998}.
In particular various shapes show up due to the interplay of 
non-linearity, fluctuation and boundary effects.
In \cite{KPZ1986}, Kardar, Parisi and Zhang proposed a 
non-linear stochastic differential equation called the KPZ equation. 
This equation, although now considered to be insufficient for 
the description of most realistic situation in nature, defines a
universality class of the surface growth phenomenon and plays 
a prominent role in the theoretical study.

In the one dimensional case we can analyze the KPZ equation exactly.
The roughness and dynamical exponents were obtained by 
the renormalization techniques~\cite{KPZ1986} 
and the Bethe Ansatz method~\cite{GS1992,Ki1995}. 
Recently we have been obtaining more sophisticated  
information for the height fluctuation
in the one-dimensional KPZ universality class; not 
only the exponents but also the height distribution itself have been obtained. 
For various probabilistic models belonging to the KPZ universality class
\cite{Jo,PS2002a,NS2004p,RS2004p,GTW,MNp},
it has turned out that the height fluctuation is equivalent 
to the Tracy-Widom distribution \cite{TW1994,TW1996}, 
the distribution of the largest 
eigenvalue in some random matrix ensemble \cite{Me1991}.


Among these models we focus on the polynuclear growth (PNG) model.
For the PNG model the relation to the random matrix theory was
first pointed out by Pr\"ahofer and Spohn~\cite{PS2000b}.  
They mapped a specific PNG model to 
the longest increasing subsequence problem 
in random permutations and then applied the
Baik-Deift-Johansson theorem~\cite{BDJ1999}.
%
%
%
The dependence of the height fluctuation 
on the geometry of the model is also studied
based on the works of Baik and Rains~\cite{BR2001a,BR2001b,BR2001c}
in random permutations with symmetries. 
As a result the deep connection with  the various  
universality classes in random matrix theory (RMT) have been revealed.
For example, for the droplet growth in an infinite line, 
the height distribution
can be described by the GUE Tracy-Widom distribution~\cite{PS2000b}
whereas in a half-infinite line, the height fluctuation at the origin  
can be described by the GSE/GOE/Gaussian  
fluctuation according to the strength of the nucleation rate at the origin
\cite{PS2000a,SI2003p}.   
On the other hand the GOE Tracy-Widom distribution represents 
the height fluctuation in 
a growth on a flat substrate~\cite{PS2000a,Ferrari2004p}.




Next we broaden our field of view from the height fluctuation at one point
to that over some region of the surface. 
In other words, we address the issue about the roughness of the surface.
The spatial configuration of the height fluctuation is expected to 
converge to the universal stochastic process after a proper scaling when
the space direction is treated as the time direction of the process.
In general a stochastic process is characterized by a dynamical correlation
function together with a fluctuation at one point.
Thus if we try to understand the universal aspect of the spatial configuration
of the fluctuation, we need information about the multi-point correlations of 
the height fluctuation.

The correlation 
between distinct points of the PNG model 
is closely related to the  multi-matrix model. 
In~\cite{PS2002b,Johansson2002p}, the multi-point equal time height 
fluctuations of the one-dimensional PNG model was studied for a 
droplet initial condition in an infinite space. 
It was found that the correlation is described by the Airy process,
which is the process of the largest eigenvalue in the Dyson's 
Brownian motion model~\cite{Dy1962} for GUE.
This also appeared in the facet fluctuation in the crystal~\cite{FS,FPS}.   
In \cite{SI2003p}, similar quantities are evaluated for a 
special value of the external source in case of droplet growth in half space.
It was shown that the correlation is described the Dyson's Brownian 
motion model which has transition between GOE/GSE to GUE. The same process 
also appeared in the vicious walk problem~\cite{NKT2003}.

In this paper we study the PNG model with external sources.
The goal of this study is to understand the universal process 
characterizing the roughness of the surface for this model.
There are mainly two different regions.
One is the region where the edge effects are dominant and the 
other is the region where the bulk dynamics prevails. 
The statistics of height fluctuation obey the one dimensional 
Brownian motion near edges and the Airy process in the bulk. 
When seen from far away, the above two types of regions are 
separated by a well distinguishable point. But we can focus into
a small region around this point, in which the edge effects and 
bulk dynamics are competing. We are especially interested in this 
region because there appear new processes describing the 
transition of the surface fluctuation.

%
For this purpose  we obtain the Fredholm determinant 
representation of the multi-point correlation function  
in the bulk region, near edges
and around the special points mentioned above.
We summarize the results as Theorems 3.1, 4.1 and 5.1. 
In particular the Fredholm determinant expressing 
the correlation in the intermediate region
appears in this paper for the first time.
They describe the transition between GOE$^2$ 
and GUE/Gaussian, or $F_0$ and Gaussian, where
GOE$^2$ means the distribution of the larger of the 
largest eigenvalues of two independent GOEs and 
$F_0$ is a certain probability distribution which has
no interpretation in RMT~\cite{BR2001a,BR2001b,BR2001c,BR2000}.





The paper is organized as follows.
In the next section, we recall the definition of the model
and some known facts are reviewed.
In section \ref{det}, we give the description of the 
equal-time multi-point fluctuation for finite system.
The asymptotic results for fixed values of parameters 
are also stated. In section \ref{goe2}, the transition 
near the GOE$^2$ fluctuation is discussed.
In the following section \ref{F0}, the transition around
$F_0$ is studied. In both section \ref{goe2} and \ref{F0} we also 
discuss the connection between Fredholm representation and 
the Riemann-Hilbert representation 
of Baik-Rains~\cite{BR2000}.
The last section is devoted to the conclusion.

\setcounter{equation}{0}
\section{Model and One-point Height Fluctuation}
\label{model}
In this article we mainly consider the 
discrete PNG model with external sources studied previously 
in \cite{BR2000,PS2000a}. First of all, we briefly explain the PNG model. 
The PNG model is a simple model of layer-by-layer growth~\cite{Me1998}. 
The discrete version of the model consists of the following three rules.
\begin{enumerate}
\item {\bf nucleation}: 
A nucleation with height $k$ is generated according to 
the geometric distribution. 
An object made by this rule is called a step. 

\item {\bf lateral growth of a step}: 
Once a step is produced, it grows laterally by one step in both directions 
during each time step. 

\item {\bf unification of steps}:
When two steps with distinct heights collide, the height in the colliding point
becomes that of the higher step.
\end{enumerate}
Note that the rule 1 is only probabilistic and the last two rules are deterministic.
These rules are illustrated in Fig.1. 

We can formulate the above rules of time evolution  mathematically
as follows. 
Let $r\in\Z$ and $t\in\N=\{0,1,2,\cdots\}$ denote the discrete 
space and time coordinates respectively and $h(r,t)$ the 
height of the surface at position $r$ and at time $t$.
The rules $1\sim3$ can be collected as 
\begin{equation}
 h(r,t+1) = \text{max}(h(r-1,t),h(r,t),h(r+1,t)) + \o(r,t+1),
\label{PNGdef}
\end{equation}
with the initial condition $h(r,0)=0$. 
Here $\o$ is the random variable expressing the height of nucleation and takes a value in $\N$. 
$\o(r,t)=0$ if $t-r$ is even or if $|r|>t$, and 
\begin{equation}
 w(i,j) = \o(i-j,i+j-1),
\end{equation}
$(i,j)\in\Z_+^2$ are geometric random variables.
The parameter of this random variable is taken to be of the form $a_ib_j$,
\begin{equation}
 \P[w(i,j) = k] = (1-a_i b_j) (a_i b_j)^k, \\  
\label{wii}
\end{equation}
for $k\in\N$. In order to consider the effect of external sources at both edges, we take
\begin{align}
\label{aPNG}
 a_j &= \begin{cases}
         \g_-, & j=1, \\ 
         \a,   & j \geq 2,
        \end{cases}
 \\
\label{bPNG}
 b_j &= \begin{cases}
         \g_+, & j=1, \\ 
         \a,   & j \geq 2.
        \end{cases}
\end{align}
The parameter $\a$ is related to the frequency of 
nucleations in the bulk whereas 
the parameter $\g_{\pm}$ represents the strength of the external 
sources at the edges. The bigger $\g_{\pm}$ is, the stronger 
the source is. In the following, we assume $\g_- > \g_+$ when we 
do some computations; the results for the case where $\g_+>\g_-$
is obtained from the symmetry.

Some snapshots of Monte Carlo simulations are given in Fig.2.
When the external sources are not very strong (Fig.(a)), 
the effects of the external sources are important 
for some region near the edges whereas
the bulk dynamics is important for the curved region at bulk.
On the other hand, when the external sources are strong (Fig.(b))
they are dominant for the whole region.
In the critical situation (Fig.(c)), they control the whole
region but a certain point. 
More precisely the shape is described as follows. Let us set
\begin{align}
 a(\b) 
 &= 
 \frac{2\a}{1-\a^2} \left( \a+\sqrt{1-\b^2} \right),
\label{adef}\\
 a_{G\pm}(\b,\g) 
 &=
  \frac{\a(1-2\a\g+\g^2)}{(\g-\a)(1-\a\g)} 
  \pm \frac{\a(\g^2-1)}{(\g-\a)(1-\a\g)} \b,
\label{agdef} 
\end{align}
and 
\begin{align}
 \b_- &=  \frac{(1-\a^2)(\g_-^2-1)}{1+\a^2-4\a\g_-+\g_-^2+\a^2 \g_-^2}, \\
 \b_+ &= -\frac{(1-\a^2)(\g_+^2-1)}{1+\a^2-4\a\g_++\g_+^2+\a^2 \g_+^2}.
\end{align}
Notice $\b_-<\b_+ \Leftrightarrow \g_+\g_-<1$.
Then the thermodynamic shape is given by the following.

\noindent
(i)
When $\b_-<\b_+$,

\begin{align}
 h(r=2\b N,t=2N)/N 
 \sim
 \begin{cases}
  a_{G-}(\b,\g_-),
  & \b < \b_- , \\
  a(\b),
  & \b_- < \b < \b_+ , \\
  a_{G+}(\b,\g_+),
  & \b > \b_+ .
 \end{cases} 
\end{align}

\noindent
(ii)
When $\b_->\b_+$,
\begin{align}
 h(r=2\b N,t=2N)/N 
 \sim
 \begin{cases}
  a_{G-}(\b,\g_-),
  & \b < \b_c , \\
  a_{G+}(\b,\g_+),
  & \b > \b_c ,
 \end{cases} 
\end{align}
with $\b_c$ being the solution of $ a_{G-}(\b,\g_-)= a_{G+}(\b,\g_+)$.

The fluctuation properties of the model 
change drastically at 
the connecting points $\b_{\pm}$ and $\b_c$ of the limiting shapes.
Let us define the two scaled height variables.
The first one is 
\begin{equation}
\label{scaledH}
 H_N(\t,\b_0) 
 = 
 \frac{h(r=2\b_0 N+2c(\b_0)N^{\frac23}\t,t=2N)
       -a(\b_0+\frac{c(\b_0)\t}{N^{1/3}}) N}
      {d(\b_0) N^{\frac13}} ,
\end{equation}
where
\begin{align}
\label{ddef}
 d(\b) &= \frac{\a^{\frac13}}{(1-\a^2)(1-\b^2)^{\frac16}}
             (\sqrt{1+\b}+\a\sqrt{1-\b})^{\frac23}
             (\sqrt{1-\b}+\a\sqrt{1+\b})^{\frac23}, \\
 c(\b) &= \a^{-\frac13} (1-\b^2)^{\frac23}
             (\sqrt{1+\b}+\a\sqrt{1-\b})^{\frac13}
             (\sqrt{1-\b}+\a\sqrt{1+\b})^{\frac13}.
\label{cdef}
\end{align}
The second is 
\begin{equation}
\label{scaledHG}
 H_N^{(G\pm)}(\b_0,\g) 
 = 
 \frac{h(r=2\b_0 N,t=2N) - a_{G\pm}(\b_0,\g) N}
      {d_G(\g) N^{\frac12}} ,
\end{equation}
where
\begin{equation}
 d_G(\g)
 =
 \frac{\sqrt{\a\g(1+\a^2-4\a\g+\g^2+\a^2\g^2)}}
      {(1-\a\g)(\g-\a)}.
\label{dgdef}
\end{equation}
About the one point height fluctuation, we have the following.
\begin{theorem}
\hspace{90pt}\\
\noindent
(i)
When $\b_- < \b_+$.

\noindent
a) For $\b_- < \b_0 < \b_+$, 
\begin{equation}
 \lim_{N\to\i} \P[H_N(0,\b_0) \leq s]
 =
 F_2(s),
\end{equation}
where $F_2$ denotes the GUE Tracy-Widom distribution, 
which is the distribution of 
the largest eigenvalue in GUE \cite{TW1994}. 

\noindent
b) For $\b_0=\b_-$ or $\b_0=\b_+$,
\begin{equation}
\label{detKlim12}
 \lim_{N\to\i} \P[H_N(0,\b_0) \leq s]
 =
 F_1(s)^2.
\end{equation}
where $F_1$ is the GOE Tracy-Widom distribution \cite{TW1996}. Thus $F_1^2$, 
which is denoted by GOE$^2$,  means the distribution of the larger of 
the largest eigenvalues in two independent GOEs.

\noindent
c) For $\b_0<\b_-$ or $\b_0>\b_+$, the fluctuation is Gaussian.
One has
\begin{equation}
\label{detKlimG}
 \lim_{N\to\i} \P[H_N^{(G\pm)}(\b_0,\g_{\pm})\leq s]
 =
 \frac{1}{\sqrt{2\pi|\b_{\pm}-\b_0|}}
 \int_{-\i}^s d\xi e^{-\frac{\xi^2}{2\pi|\b_{\pm}-\b_0|}}.
\end{equation}

\noindent
(ii)
When $\b_- > \b_+$.

\noindent
a) For $\b_0<\b_c$ or $\b_0>\b_c$, the fluctuation is Gaussian (cf. (i-c)).

\noindent
b) For $\b_0=\b_c$, the fluctuation might be given by
\begin{align}
 \lim_{N\to\i} \P[H_N^{(G_-)}(\b_0,\g_{-})\leq s]
 =\int_{-\i}^{s}d\xi_1\frac{e^{-\frac{\xi_1^2}{2(\b_--\b_0)}}}
{\sqrt{2\pi(\b_--\b_0)}}\int_{-\i}^{s}d\xi_2
\frac{e^{-\frac{\xi_2^2}{2(\b_0-\b_+)}
\frac{d_G(\g_-)^2}{d_G(\g_+)^2}}}
{\sqrt{2\pi(\b_0-\b_+)}\frac{d_G(\g_+)}{d_G(\g_-)}}.
\end{align}

\noindent

\noindent
(iii)
When $\b_- = \b_+$.

\noindent
a) For $\b_0<\b_-$ or $\b_0>\b_-$, the fluctuation is Gaussian (cf. (i-c)).

\noindent
b) For $\b_0=\b_-$, 
\begin{equation}
\label{detKlim0}
 \lim_{N\to\i} \P[H_N(0,\b_0) \leq s]
 =
 F_0(s).
\end{equation}
Here $F_0$ is a certain distribution with mean zero explained in 
~\cite{BR2001b,BR2001c,BR2000}.
\end{theorem}

These are obtained as corollaries of the theorem in the next section.
The special case of the height at the origin ($\b_0=0$) was previously 
studied in section 4 of \cite{BR2000} using the connection 
of the problem to the combinatorics of Young tableaux.
The limiting distribution was obtained using the Riemann-Hilbert 
method and the results were given in terms of the solution to the 
Painlev\'e equation.
In \cite{PS2000a}, for the continuous model, the basic picture 
of this theorem was expected based on a physical argument 
but has not been shown explicitly.
These distributions also appear in the fluctuation properties of
the one-dimensional asymmetric simple exclusion process
(ASEP)~\cite{PS2002a,NS2004p}.

\setcounter{equation}{0}
\section{Multi-point Height Fluctuation}
\label{det}
As was observed in \cite{Johansson2002,PS2002b,Johansson2002p,SI2003p},
the equal time multi-point correlation of the PNG model can be 
analyzed by extending the original model to the multi-layer version. 
The weight of the multi-layer version is equivalent to that of 
non-intersecting many-body random walk.  
In particular, for the model under consideration, 
this can be borrowed from the results in 
\cite{Johansson2002p}.
Following \cite{Johansson2002p}, we only consider an odd time $M=2N-1$
in the sequel.
Let us consider the weight for the configuration $\{x_j^r\}\equiv\bar{x}$ 
of $n$ non-intersecting paths from the time $r=-M$ to $r=M$ given by
\begin{equation}
\label{WnM}
 w_{n,M}(\bar{x}) 
 = 
 \prod_{r=-M}^{M-1} \det (\p_{r,r+1}(x_i^r,x_j^{r+1}))_{i,j=1}^n,
\end{equation}
where
\begin{align}
 \p_{2j-1,2j}(x,y) 
 &=
 \begin{cases}
  (1-a_{j+N}) a_{j+N}^{y-x}, & y \geq x, \\
  0, & y<x,
 \end{cases}
 \\
 \p_{2j,2j+1}(x,y) 
 &=
 \begin{cases}
  0, & y>x, \\
  (1-b_{N-j}) b_{N-j}^{x-y}, & y \leq x, 
 \end{cases}
\end{align}
and $x_i^M=x_i^{-M}=1-i$ ($i=1,2,\cdots,n$) is fixed.
Note that the same weight gives a weight for a 
time evolution by the time 
$M$ of the PNG model. 
For the PNG model with external sources we consider 
in this paper, the parameters $a_j,b_j$'s are taken to be
(\ref{aPNG}), (\ref{bPNG}).
Strictly speaking, the weight of the multi-layer PNG model and 
the weight (\ref{WnM}) with (\ref{aPNG}),(\ref{bPNG}) and $n=N$ 
are slightly different. Essentially the same remark was 
already given in \cite{SI2003p}.
The difference is, however, negligible in the scaling 
limit in which we are mainly interested in this paper. 

For each fixed $\g_{\pm}$, as $N\to\i$, we have the following 
results.
\begin{theorem}
\hspace{90pt}\\
\noindent
(i) 
When $\b_- < \b_+$.

\noindent
a)For $\b_- < \b_0 < \b_+$, the equal time multi-point distribution function
is described by the following Fredholm determinant. 
\begin{align}
\label{detKlim_o0}
 &\quad\lim_{N\to\i} \P[H_N(\t_1,\b_0) \leq s_1, 
   \cdots , H_N(\t_m,\b_0) \leq s_m] \notag\\
 &= \sum_{k=0}^{\i} \frac{1}{k!} 
 \sum_{n_1=1}^m \int d\xi_1 \cdots 
 \sum_{n_k=1}^m \int d\xi_k 
 ~\G(\t_{n_1},\xi_1) \cdots \G(\t_{n_k},\xi_k)  
 \det(\K(\t_{n_l},\xi_l; \t_{n_{l'}},\xi_{l'}))_{l,l'=1}^k,\notag\\
 &\equiv\det(1+\K \G),
\end{align}
where $\G(\t_j,\xi)=-\chi_{(s_j,\i)}(\xi)$ ($j=1,2,\cdots,m$).($\chi_J$ is the 
characteristic function.)
The kernel $\K$ is the extended Airy kernel,
\begin{equation}
 \label{K2def}
 \K_2(\t_1,\xi_1;\t_2,\xi_2)
 =
 \begin{cases}
  \int_0^{\i} d\l e^{-\l(\t_1-\t_2)} \Ai(\xi_1+\l) \Ai(\xi_2+\l), 
  & \t_1 \geq \t_2 ,\\
  -\int_{-\i}^0 d\l e^{-\l(\t_1-\t_2)} \Ai(\xi_1+\l) \Ai(\xi_2+\l), 
  & \t_1 < \t_2 .
 \end{cases}
\end{equation}

\noindent
b)
For $\b_0=\b_-$ or $\b_0=\b_+$, we have (\ref{detKlim_o0})
with a different kernel. It is denoted as $\K_{12}$ and is given by
\begin{align}
 &\quad 
 \K_{12}(\t_1,\xi_1;\t_2,\xi_2) \notag\\
 &=
 \begin{cases}
  \K_2(\t_1,\xi_1;\t_2,\xi_2)
  +\Ai(\xi_1) \int_0^{\i} d\l e^{- \t_2 \l}\Ai(\xi_2-\l) ,
  & \t_2>0,    \\
  \K_2(\t_1,\xi_1;\t_2,\xi_2)
  -\Ai(\xi_1) \int_0^{\i} d\l e^{\t_2\l}\Ai(\xi_2+\l)   
  +\Ai(\xi_1) e^{\frac{\t_2^3}{3}-\xi_2\t_2},
  & \t_2<0 .
 \end{cases}
\end{align}

\noindent
c)
In the region where $\b_0<\b_-$ or $\b_0>\b_+$, the fluctuation is 
equivalent to those of the Brownian motion.
In terms of the Fredholm representation, 
when $\b_1 < \b_2 <\cdots < \b_m<\b_-$, we have
\begin{equation}
 \lim_{N\to\i} \P[H_N^{(G-)}(\b_1,\g_-) \leq s_1, \cdots , 
                  H_N^{(G-)}(\b_m,\g_-) \leq s_m]
 =
 \det(1+\K \G),
\label{gausuasy}
\end{equation}
where the kernel is 
\begin{equation}
 \K_{G-}(\b_1, \xi_1;\b_2, \xi_2) 
 =
 \begin{cases}
  \frac{e^{-\frac{\xi_1^2}{2(\b_--\b_1)}}}{\sqrt{2\pi(\b_--\b_1)}},
  & \b_1 \geq \b_2, \\
  \frac{e^{-\frac{\xi_1^2}{2(\b_--\b_1)}}}{\sqrt{2\pi(\b_--\b_1)}}
  -
  \frac{e^{-\frac{(\xi_2-\xi_1)^2}{2(\b_2-\b_1)}}}{\sqrt{2\pi(\b_2-\b_1)}},
  & \b_1 < \b_2 .
 \end{cases}
\label{kernelG}
\end{equation}

The results for the case where $\b_+<\b_1<\b_2<\cdots <\b_m$
is analogous.

\noindent
(ii) 
When $\b_->\b_+$.

\noindent
a)For $\b<\b_c$, the same fluctuation as~\eqref{kernelG} is obtained.   

\noindent
(iii)
When $\b_- = \b_+$.

\noindent
a)For $\b<\b_-$ and $\b>\b_-$, 
the same fluctuation as~\eqref{kernelG} is obtained. 

\noindent
b)
For $\b = \b_c$, the results can be obtained as a limiting case
of Theorem 5.1 in section \ref{F0}.
\end{theorem}

\vspace{3mm}\noindent
{\bf Remarks.} 
\begin{enumerate}
\item The process characterized by the Fredholm determinant with
the extended Airy kernel~\cite{FNH1999,Mac1994} in (i-a) 
is called the Airy process \cite{PS2002b,Johansson2002p}.
This is the same as the process of the largest 
eigenvalue in the Dyson's Brownian Motion model~\cite{Dy1962}
between unitary ensembles.

\item For (i-c) we can easily calculate the multi-point joint 
distributions. For instance the two-point joint distribution is 
calculated as
  \begin{align}
    & \lim_{N\to\i} \P[H_N^{(G-)}(\b_1,\g_-) \leq s_1 , 
                  H_N^{(G-)}(\b_2,\g_-) \leq s_2] \notag\\
    &= 1-\int_{s_1}^{\infty}d \xi_1 \K_{G_-}(\b_1,\xi_1;\b_1,\xi_1)
    -\int_{s_2}^{\infty}d \xi_2 \K_{G_-}(\b_2,\xi_2;\b_2,\xi_2)
    \notag\\
    &\quad+\frac12\int_{s_1}^{\infty}d\xi_1\int_{s_2}^{\infty}d\xi_2
    \left|
      \begin{array}{@{\,}cc@{\,}}
         \K_{G_-}(\b_1,\xi_1;\b_1,\xi_1)& \K_{G_-}(\b_1,\xi_1;\b_2,\xi_2) \\
         \K_{G_-}(\b_2,\xi_2;\b_1,\xi_1)& \K_{G_-}(\b_2,\xi_2;\b_2,\xi_2)
      \end{array}
    \right| \notag\\
    &\quad+\frac12\int_{s_1}^{\infty}d\xi_1\int_{s_2}^{\infty}d\xi_2
    \left|
      \begin{array}{@{\,}cc@{\,}}
         \K_{G_-}(\b_2,\xi_2;\b_2,\xi_2)& \K_{G_-}(\b_2,\xi_2;\b_1,\xi_1) \\
         \K_{G_-}(\b_1,\xi_1;\b_2,\xi_2)& \K_{G_-}(\b_1,\xi_1;\b_1,\xi_1)
      \end{array}
    \right| \notag\\
   &=\int_{-\infty}^{s_1}d\xi_1\int_{-\infty}^{s_2}d\xi_2
  \frac{e^{\frac{-(\xi_2-\xi_1)^2}{2(\b_2-\b_1)}}}{\sqrt{2\pi(\b_2-\b_1)}}
  \frac{e^{\frac{-\xi_2^2}{2(\b_--\b_2)}}}{\sqrt{2\pi(\b_--\b_2)}}.
  \label{brown}
  \end{align}
Note that the 3 by 3 determinant, 
$\det[K_{G_-}(t_i,\xi_i;t_j,\xi_j)]_{i,j=1}^3$ becomes zero when
$t_j=\b_1, \b_2$ since the 2 by 2 determinant at each $\b_j$, 
$\det[K_{G_-}(\b_k,\xi_i;\b_k,\xi_j)]_{i,j=1}^3~~(k=1,2)$, vanishes.
Hence
bigger determinants also vanish. The integrand in~\eqref{brown} 
represents the propagation of a Brownian particle.

\end{enumerate}

\medskip
\noindent
{\bf Proof.} 
Here we prove (i-a), (i-c), (ii-a) and (iii-a). 
The remainings, (i-b) and (iii-b), will be obtained as 
corollaries of the results in the following sections.

First we start our proof with the fact that  
equal time multi-point correlation of height fluctuations 
at the odd time $M=2N-1$ has the Fredholm
representation~\cite{Johansson2002p},
\begin{align}
& \P\left[h(r_1,M)\le l_1,h(r_2,M),\le l_2\cdots ,h(r_m,M)\le l_m\right]
\notag\\
&=\sum_{k=0}^{\infty}\frac{1}{k!}\sum_{i_1=1}^{m}\sum_{x_{1}}\cdots
\sum_{i_k=1}^{m}\sum_{x_{k}}g(r_{i_1},x_{1})\cdots g(r_{i_k},x_{k})
\det(K_N(r_{i_l},x_{l};r_{i_{l'}},x_{l'}))_{l,l'=1}^{k}\notag\\
&\equiv\det(1+K_N g),
\label{Fdet}
\end{align}
where 
\begin{align}
&g(r_j,x_i)
=-\chi_{(l_j,\infty)}(x_i), \notag\\
 &
K_N(r_1=2u_1,x_1;r_2=2u_2,x_2)=\tilde{K}_N(2u_1,x_1;2u_2,x_2)
-\phi_{2u_1,2u_2}(x_1,x_2), 
\label{dif2}\\
&\tilde{K}_N(2u_1,x_1;2u_2,x_2), \notag\\
&=
 \frac{ (1-\a)^{2(u_2-u_1)}}{(2\pi i)^2}
 \int_{C_{R_1}} \frac{dz_1}{z_1}\int_{C_{R_2}}\frac{dz_2}{z_2}
 \frac{z_2^{x_2}}{z_1^{x_1}} \frac{z_1}{z_1-z_2}
 \frac{(1-\a/z_1)^{N-1+u_1}(1-\a z_2)^{N-1-u_2}}
      {(1-\a z_1)^{N-1-u_1}(1-\a/z_2)^{N-1+u_2}}
 \notag\\
 &\quad \times
 \frac{1-\g_-/z_1}{1-\g_-/z_2} \frac{1-\g_+ z_2}{1-\g_+ z_1}, 
\label{dif}\\
&\p_{r_1,r_2}(x_1,x_2) 
=
\begin{cases}
 \frac{(1-\a)^{2(u_2-u_1)}}{2\pi i}
 \int_{C_1} \frac{dz}{z} 
 z^{x_2-x_1} [(1-\a z)(1-\a/z)]^{u_2-u_1}, & u_2> u_1, \\
 0,   & u_2\leqslant u_1 .
\end{cases}
\label{psi}
\end{align}
In ~(\ref{dif}), 
$C_{R_i}$ means a contour with a radius $R_i$ and $C_1$ is the unit circle.
In both cases, they enclose the origin anticlockwise.   
One takes the radiuses $C_{R_1}, C_{R_2}$
in a way that $\g_{-}<C_{R_2}<C_{R_1}<1/{\g_+}$.
Note that~\eqref{Fdet} is valid for $\g_+\g_-<1$.
    
Next we discuss the asymptotics by applying the saddle point method 
to the kernel~(\ref{dif}) and~(\ref{psi}).

\begin{center}
\bf{proof of (i-a)}  
 \end{center}
We prove \eqref{detKlim_o0} and \eqref{K2def} along the same line as the derivation 
of the Proposition 4.1
in \cite{SI2003p}.
We  give only the outline of the proof. 
\begin{flushleft}
 \bf{$\bullet $ Asymptotics of $\tilde{K}_N$ }
\end{flushleft}
We set
\begin{align}
 \label{Sa}
 &\quad
 \tilde{K}_N(r_1=2u_1,x_1;r_2=2u_2,x_2) \notag\\
 &=
 \frac{(1-\a)^{2(u_2-u_1)}}{(2\pi i)^2}
 \int_{C_{R_1}} \frac{dz_1}{z_1} \int_{C_{R_2}} \frac{dz_2}{z_2}
 \frac{z_2^{x_2-N(\mu_2-1)}}{z_1^{x_1-N(\mu_1-1)}}
 e^{N(g_{\mu_1,\b_1}(z_1)+g_{\mu_2,\b_2}(1/z_2))} \notag\\
 &\quad\times
 \frac{z_1}{z_1-z_2}
 \frac{1-\g_-/z_1}{1-\g_-/z_2} \frac{1-\g_+ z_2}{1-\g_+ z_1},
\end{align}
where $\b_1=u_1/N,\b_2=-u_2/N$,  $\mu_1,\mu_2$ are arbitrary
constants at this stage and 
\begin{equation}
\label{defg}
g_{\mu,\b}(z)
=
(1+\b)\log(z-\a)-(1-\b)\log(1-\a z)-(\mu+\b)\log z.
\end{equation} 
Using $a(\b), c(\b), d(\b)$ defined in 
\eqref{adef},~\eqref{ddef} and~\eqref{cdef}
respectively, we consider the scaling,
\begin{align}
&\b_i=\b_0+\frac{c(\b_0)}{N^{\frac13}}\t_i,\\
&x_i=a(\b_i)N+d(\b_0)N^{\frac13}\xi_i.  
\end{align}
When we fix $\mu_c(\b)=a(\b)+1$, 
two critical points of $g_{\mu,\b}(z)$ are combined to the  
double critical point $p_c(\b)$ given by 
\begin{equation}
 \label{pcdef}
 p_c(\b)
 =
 p(\mu_c(\b),\b)
 =
 \frac{\sqrt{1+\b}+\a \sqrt{1-\b}}{\sqrt{1-\b}+\a\sqrt{1+\b}}, 
\end{equation} 
 where 
$g'_{\mu_c(\b),\b}(p_c(\b)) = g''_{\mu_c(\b),\b}(p_c(\b))=0$.
Since $\g_-<p_c(\b)<\frac1{\g_+}$ for $\a<\g<\frac1\a$, $\b_-<\b<\b_+$, 
we can deform the contour of $z_i$ to
\begin{align}
 \label{pathz1}
 z_1           &= p_c(\b_1)\left(1-\frac{i}{d(\b_0)N^{1/3}}w_1\right)
\sim p_c(\b_0)\left(1 + \frac{1}{d(\b_0)N^{1/3}}(\t_1-iw_1)\right), \\
 \label{pathz2}
 \frac{1}{z_2} &= p_c(\b_2)\left(1-\frac{i}{d(\b_0)N^{1/3}}w_2\right)
\sim  \frac{1}{p_c(\b_0)}\left(1 - \frac{1}{d(\b_0)N^{1/3}}(\t_2+iw_2)\right).
\end{align}
Then we find 
\begin{align}
 \label{Gb1}
 N g_{\mu_c(\b_1),\b_1}(z_1)
 &\sim
 N\l_0(\b_0)+ \l_1(\b_0)c(\b_0)N^{2/3}\t_1 \notag\\
 &\quad +\l_2(\b_0)c(\b_0)^2 N^{1/3}\t_1^2 +\l_3(\b_0)c(\b_0)^3 \t_1^3
 +\frac{i}{3}w_1^3, \\
 \label{Gb2}
 N g_{\mu_c(\b_2),\b_2}(\frac{1}{z_2})
 &\sim
 -N \l_0(\b_0)-\l_1(\b_0)c(\b_0)N^{2/3}\t_2 \notag\\
 &\quad
 -\l_2(\b_0)c(\b_0)^2 N^{1/3}\t_2^2 -\l_3(\b_0)c(\b_0)^3 \t_2^3
+\frac{i}{3}w_2^3, 
\end{align}
where 
\begin{equation}
\label{deflam}
  \left.\l_i(\b_0)=\frac{d^i}{d\b^i}g_{\mu_c(\b),\b}(p_c(\b))\right|_{\b=\b_0}.
\end{equation}
Similarly one gets
\begin{align}
& 
\label{z1z2asy2}
\frac{z_1}{z_1-z_2}
 \sim
 -\frac{d(\b_0)N^{1/3}}{\t_2-\t_1+i(w_1+w_2)},
\\
&
\frac{z_2^{x_2+N(1-\mu_c(\b_2))}}{z_1^{x_1+N(1-\mu_c(\b_1))}}
 \sim
 (p_c(\b_0))^{(\xi_2-\xi_1)d(\b_0) N^{1/3}}
 e^{\xi_2 \t_2-\xi_1 \t_1+i\xi_1 w_1+i\xi_2 w_2}
.
\label{z1z2asy}
\end{align}
Substituting \eqref{pathz1}--~\eqref{z1z2asy} into~\eqref{Sa},
one obtains
\begin{align}
\label{Sasy}
 \tilde{K}_N
 &\sim
 (1-\a)^{2(u_2-u_1)} (p_c(\b_0))^{(\xi_2-\xi_1)d(\b_0)N^{1/3}}
 \frac{1}{d(\b_0)N^{1/3}} \notag\\
 &\quad
 e^{ \l_1(\b_0)c(\b_0)N^{2/3}(\t_1-\t_2)
    +\l_2(\b_0)c(\b_0)^2 N^{1/3}(\t_1^2-\t_2^2)
    +\l_3(\b_0)c(\b_0)^3 (\t_1^3-\t_2^3)+\xi_2 \t_2-\xi_1 \t_1} \notag\\
 &\quad
 \frac{1}{4\pi^2}\int_{\text{Im}w_1=\eta_1} dw_1 
                 \int_{\text{Im}w_2=\eta_2} dw_2 
 \left(-\frac{1}{\tau_2-\tau_1+i(w_1+w_2)}\right)
 e^{i\xi_1 w_1 + i\xi_2 w_2 + \frac{i}{3}(w_1^3+w_2^3)}\\
 &=
(1-\a)^{2(u_2-u_1)} (p_c(\b_0))^{(\xi_2-\xi_1)d(\b_0)N^{1/3}}
 \frac{1}{d(\b_0)N^{1/3}} \notag\\
 &\quad
 e^{ \l_1(\b_0)c(\b_0)N^{2/3}(\t_1-\t_2)
    +\l_2(\b_0)c(\b_0)^2 N^{1/3}(\t_1^2-\t_2^2)
    +\l_3(\b_0)c(\b_0)^3 (\t_1^3-\t_2^3)+\xi_2 \t_2-\xi_1 \t_1} \notag\\
 &\quad
 \int_0^{\i} d\l e^{-\l(\t_1-\t_2)} \Ai(\xi_1+\l) \Ai(\xi_2+\l),
\label{KNresult}
\end{align}
where $\eta_i>0$ is a convergence factor.
\begin{flushleft}
 \bf{$\bullet$ Asymptotics of $\phi_{2u_1,2u_2}(x_1,x_2)$}
\end{flushleft}

Next we consider the asymptotics of $\phi_{2u_1,2u_2}(x_1,x_2)$.
For $u_1<u_2$ 
\begin{align}
 &\quad
 \p_{r_1,r_2}(x_1,x_2) \notag\\
 &=
 \frac{(1-\a)^{2(u_2-u_1)}}{2\pi i}
 \int_{C_1} \frac{dz}{z}
 z^{x_2-x_1} [(1-\a z)(1-\a/z)]^{u_2-u_1} \notag\\
 &=
 \frac{(1-\a)^{2(u_2-u_1)}}{2\pi i}
 \int_{C_1} \frac{dz}{z}
 z^{x_2-N(\mu_c(\b_2)-1)-x_1+N(\mu_c(\b_1)-1) }
 e^{N g_{\mu_c(\b_1),\b_1}(z) + N g_{\mu_c(\b_2),\b_2}(1/z)}.
\end{align}
We set 
\begin{align}
 z = p_c(\b_0)\left(1+\frac{i\sigma}{d(\b_0) N^{1/3}}\right) 
   &\sim p_c(\b_1)\left(1-\frac{1}{d(\b_0)N^{1/3}}(\t_1-i\sigma)  \right)
 \label{e1}\\ \label{e2}
   &\sim\frac{1}{p_c(\b_2)}
\left(1-\frac{1}{d(\b_0)N^{1/3}}(\t_2-i\sigma)\right).
\end{align}
Applying these to $g_{\mu_c(\b_1),\b_1}(z)$
and $g_{\mu_c(\b_2),\b_2}(1/z)$ respectively, we get 
\begin{align}
&   N g_{\mu_c(\b_1),\b_1}(z)
 \sim
 N g_{\mu_c(\b_1),\b_1}\big(p_c(\b_1)\big) -\frac13 (\t_1-i\sigma)^3, \\
& N g_{\mu_c(\b_2),\b_2}\left(\frac{1}{z}\right)
 \sim
 N g_{\mu_c(\b_2),\b_2}\big(p_c(\b_2)\big) + \frac13 (\t_2-i\sigma)^3. 
\end{align}
One also obtains
\begin{equation}
 z^{x_2-N(\mu_c(\b_2)-1)-x_1+N(\mu_c(\b_1)-1)} 
 \sim
 \big(p_c(\b_0)\big)^{d(\b_0)N^{1/3}(\xi_2-\xi_1)}
 e^{i\sigma (\xi_2-\xi_1)}.
\end{equation}
Hence one finds
\begin{align}
 \p_{r_1,r_2}(x_1,x_2)
 &\sim
 (1-\a)^{2(u_2-u_1)} \big(p_c(\b_0)\big)^{(\xi_2-\xi_1)d(\b_0)N^{1/3}}
 \frac{1}{d(\b_0)N^{1/3}} \notag\\
 &\quad
 e^{ \l_1(\b_0)c(\b_0)N^{2/3}(\t_1-\t_2)
    +\l_2(\b_0)c(\b_0)^2 N^{1/3}(\t_1^2-\t_2^2)
    +\l_3(\b_0)c(\b_0)^3 (\t_1^3-\t_2^3)-\frac{\t_1^3}{3}+\frac{\t_2^3}{3}} 
 \notag\\
 &\quad
 \frac{1}{2\pi}\int_{-\i}^{\i} d\sigma
 e^{i(\xi_2-\xi_1+\t_1^2-\t_2^2)\sigma-(\t_2-\t_1)\sigma^2} 
\label{pasy}\\
&=
 (1-\a)^{2(u_2-u_1)} \big(p_c(\b_0)\big)^{(\xi_2-\xi_1)d(\b_0)N^{1/3}}
 \frac{1}{d(\b_0)N^{1/3}} \notag\\
 &\quad
 e^{ \l_1(\b_0)c(\b_0)N^{2/3}(\t_1-\t_2)
    +\l_2(\b_0)c(\b_0)^2 N^{1/3}(\t_1^2-\t_2^2)
    +\l_3(\b_0)c(\b_0)^3 (\t_1^3-\t_2^3)-\frac{\t_1^3}{3}+\frac{\t_2^3}{3}} 
 \notag\\
 &~~\int_{-\i}^{\i} d\l e^{-\l(\t_1-\t_2)} \Ai(\xi_1+\l) \Ai(\xi_2+\l).
\label{phiresult}
\end{align} 

From~\eqref{KNresult} and~\eqref{phiresult} we get the desired expression. 
Note that the pre-factor of the 
extended Airy kernel  does not affect the Fredholm determinant. 

\begin{center}
  {\bf proof of (i-c),(ii-a) and (iii-a)}
\end{center}
One can prove (i-c) in a fairly similar manner to 
(i-a) but an essential modification 
is necessary for the proof of the other two cases.
This results from the fact that 
the model defined by~\eqref{PNGdef}--~\eqref{bPNG} is 
not well-defined in the latter case since the parameter of 
geometric distribution at $i=1,j=1$, which is equal to 
$\g_+\g_-$, is greater than 1.
Thus we consider a slightly modified model in which the parameter of the 
random variable at the point (1,1) is zero as in~\cite{BR2000}. 
When $\g_+\g_-<1$, the modification is unnecessary but we consider the 
modified model here because the modification does not change 
the asymptotic properties of the model and allows us to treat all 
cases in a parallel fashion. 

For $~\g_+\g_-<1$ we can get easily the relation of the 
equal time multi-point correlation function  
between these models by generalizing the relation in one-point 
case~\cite{BR2000}.  
Let $h^+(r_i,t)$ represent the height in the original model
and $h(r_i,t)$ the height of the modified model. 
One has   
\begin{align}
&\quad\P[h(r_1,t=M\equiv 2N-1)<l_1,\cdots ,h(r_m,t=M)<l_m]\notag\\ 
&=\frac{\P[h^+(r_1,M)<l_1,\cdots ,h^+(r_m,M)<l_m] 
-\g_+\g_-\P[h^+(r_1,M)<l_1-1,\cdots ,h^+(r_m,M)<l_m-1]}{1-\g_+\g_-}.
\label{ModDef}
\end{align}
When $\g_+\g_-<1$,
$\P[h^+(r_1,M)<l_1,\cdots ,h^+(r_m,M)<l_m]$ is represented as 
the Fredholm determinant~\eqref{Fdet} 
with the kernel~\eqref{dif2}--~\eqref{psi}.
The problem is that the Fredholm determinant is not well-defined
for $\g_+\g_->1$ (See (3.45) below).
We would like to obtain another representation applicable
for the case where $\g_+\g_->1$ by modifying the Fredholm representation.

We start from the 
Fredholm representation~\eqref{Fdet} for $\g_+\g_-<1$. Using the relation
\begin{equation}
\frac{z_1}{z_1-z_2}\frac{1-\g_+z_2}{1-\g_+z_1}
\frac{1-\g_-/z_1}{1-\g_-/z_2}
=
\frac{z_2}{z_1-z_2}+\frac{1-\g_+\g_-}{(1-\g_+z_1)(1-\g_-/z_2)},
\end{equation}
the kernel $\tilde{K}_N$ can be divided into two terms,
\begin{align}
 &\quad 
 \tilde{K}_N(2u_1,x_1;2u_2,x_2) \notag\\
 &=\frac{1}{(2\pi i)^2}
 \int_{C_{R_1}} \frac{dz_1}{z_1}\int_{C_{R_2}}\frac{dz_2}{z_2}
 \frac{z_2^{x_2}}{z_1^{x_1}}\frac{z_2}{z_1-z_2}
 \frac{F(u_1,z_1)}{F(u_2,z_2)}\notag\\
 &+(1-\g_+\g_-)\frac1{2\pi i} \int_{C_{R_1}}\frac{dz_1}{z_1}
 \frac1{z_1^{x_1}}F(u_1,z_1)
 \frac1{1-\g_+z_1}\times\frac1{2\pi i}
 \int_{C_{R_2}}\frac{dz_2}{z_2}\frac{z_2^{x_2}}{F(u_2,z_2)}
 \frac1{1-\g_-/z_2}\notag\\
 &\equiv \tilde{K}_{2N}(u_1,x_1;u_2,x_2)+(1-\g_+\g_-)D_+(u_1,x_1)D_-(u_2,x_2),
\end{align}
where
\begin{equation}
  F(u,z)=\frac{(1-\a/z)^{N-1+u}}{(1-\a)^{2u}(1-\a z)^{N-1-u}}.
\end{equation}
Remember that this expression was derived under the 
condition $\a<\g_-<R_2<R_1<1/\g_+<1/\a$.
The Fredholm determinant can be deformed to
\begin{align}
&\quad \det\left[1+K_Ng\right] \notag\\
&=
\det\left[1+ K_{2N}g\right]\left\{1-(1-\g_+\g_-)
\sum_{k,k'=1}^{m}\sum_{x_1=l_k}^{\i}
\sum_{x_2=l_{k'}}^{\i}D_-(u_k,x_2)
E(u_k,x_1;u_{k'},x_2)D_+(u_{k'},x_2)\right\},
\label{DefoKn}
\end{align}  
where
\begin{align}
 &K_{2N}(u_1,x_1;u_2,x_2)
 =\tilde{K}_{2N}(u_1,x_1;u_2,x_2)- \phi_{2u_1,2u_2}(x_1,x_2) ,
 \notag\\
 &E(u_1,x_1;u_2,x_2)=(1-K_{2N})^{-1}(u_1,x_1;u_2,x_2).
\end{align}
In the above equation let us focus our attention to the contribution of 
terms,
\begin{equation}
  (1-\g_+\g_-)\sum_{k=1}^{m}\sum_{x=l_k}^{\i}D_-(u_k,x)D_+(u_k,x),
\label{Delpar}
\end{equation}
which arises from the 'delta function' part of $E(u_1,x_1;u_2,y_1)$. 
Here one divides $D_{\pm}$ into two parts,
\begin{align}
 &D_-(u,x)=\frac1{2\pi i}\int_{C_{R'_2}}\frac{dz_2}{z_2}\frac{z_2^x}{F(u,z_2)}
\frac{1}{1-\g_-/z_2}
         +\frac{\g_-^x}{F(u,\g_-)}\equiv D_{1-}(u,x)+D_{2-}(u,x), \notag\\
 &D_+(u,x)=\frac1{2\pi i}\int_{C_{R'_1}}\frac{dz_1}{z_1}\frac1{z_1^x}
   \frac{F(u,z_1)}{1-\g_+z_1}
          +\g_+^x F(u,1/\g_+)\equiv D_{1+}(u,x)+D_{2+}(u,x),
\end{align}
where the radiuses of contours $R'_i$ are
taken to satisfy $\a<R'_2<\g_-<1/\g_+<R'_1<1/\a$.
Notice the second terms appear from the contribution of
the poles at $\g_-$ in $D_-E(u,x)$ and at $1/\g_+$ in $D_+(k,x)$.
One finds
\begin{align}
&\quad
(1-\g_+\g_-)\sum_{x=l_k}^{\i}D_+(u_k,x)D_-(u_k,x) \notag\\
&=
(1-\g_+\g_-)\sum_{k=1}^m\sum_{x=l_k}
\Big[ D_{1+}(u_k,x)D_{1-}(u_k,x)
     +D_{1+}(u_k,x)D_{2-}(u_k,x) \notag\\
&\quad
     +D_{2+}(u_k,x)D_{1-}(u_k,x)
     +D_{2+}(u_k,x)D_{2-}(u_k,x)
\Big]. 
\label{AnCo}
\end{align}
The last term can be rewritten as
\begin{equation}
 (1-\g_+\g_-)\sum_{x=l_k}^{\i}D_{2+}(u_k,x)D_{2-}(u_k,x) 
=(1-\g_+\g_-)\sum_{x=l_k}^{\i}(\g_+\g_-)^x \frac{F(u_k,1/\g_+)}{F(u_k,\g_-)}
=(\g_+\g_-)^{l_k}\frac{F(u_k,1/\g_+)}{F(u_k,\g_-)}.
\end{equation}
One notices that the series on the middle is divergent when
$\g_+\g_->1$ but that the difficulty is avoided in the right most 
expression. In addition one can easily find that such a difficulty
does not arise for the other terms in~\eqref{AnCo} and the remaining 
terms of 
$(1-\g_+\g_-)\sum D_-(u_k,x_1)E(u_k,x_1;u_{k'},x_2)D_+(u_{k'},x_2)$ 
in~\eqref{DefoKn} with the contribution of
\eqref{Delpar} subtracted,
\begin{align}
\sum_{k,k'=1}^m\sum_{x_1=l_k}^{\i}\sum_{x_2=l_{k'}}^{\i}D_-(u_k,x_1)
E'(u_k,x_1;u_{k'},x_2)D_+(u_{k'},x_2),
\end{align}
where
\begin{align}
E'(u_k,x_1;u_{k'},x_2)=
\begin{cases}
E(u_{k},x_1;u_{k'},x_2)-\delta^{+}_k(x_1,x_2),& \text{for~~} k=k',\\
E(u_{k},x_1;u_{k'},x_2), & \text{for~~} k\neq k',
\end{cases}
\end{align}
and 
\begin{equation}
\sum_{y=l_k}^{\i}\delta^+_k(x,y)f(y)=f(x).
\end{equation}
Eventually one obtains the representation of 
$\P[h^+(r_1,M)<l_1,\cdots ,h^+(r_m,M)<l_m]$,
\begin{align}
&\P[h^+(r_1,M)<l_1,\cdots ,h^+(r_m,M)<l_m] \notag\\
&=\det\left[1+ K_{2N}g\right]
\left(1-\sum_{k=1}^m(\g_+\g_-)^{l_k}\frac{F(u_k,1/\g_+)}{F(u_k,\g_-)}
\right.\notag\\&\quad -(1-\g_+\g_-)
\left\{\sum_{k,k'=1}^{m}
\sum_{x_1=l_k}^{\i}\sum_{x_2=l_{k'}}^{\i}
D_-(u_k,x_1)E'(u_k,x_1;u_{k'},x_2)D_+(u_{k'},x_2)\right.
\notag\\ 
&\left.\left.\quad+\sum_{k=1}^m
\sum_{x=l_k}\Big[D_{1+}(u_k,x)D_{1-}(u_k,x)
+D_{1+}(u_k,x)D_{2-}(u_k,x)+
D_{2+}(u_k,x)
D_{1-}(u_k,x)\Big] \right\}\right),
\label{NewRep}
\end{align}
which is well-defined for $\g_+\g_->1$ as well.
Note that, when $\g_+\g_->1$, the original meaning as a probability
is lost in~\eqref{NewRep}.

Next we consider the asymptotics of~\eqref{NewRep}. We set
$u_i=\b_iN, l_i=a_{G_-}(\b_i,\g_-)N + d_G(\g_-)N^{\frac12}s_i$\\
$\left(x_i=a_{G_-}(\b_i,\g_-)N + d_G(\g_-)N^{\frac12}\xi_i \right)$
and $N\rightarrow\i$. 
Let us notice 
\begin{equation}
 D_{1+}(u,x) 
 = 
 \frac{(1-\a)^{-2u}}{2\pi i} \int_{C_{R_1'}}\frac{d z}{z}
 \frac{1-\a z}{(1-\a/z)(1-\g_+ z)} e^{Ng_{y+1,\b}(z)},     
\end{equation}
where $u=N\b$ and $x=Ny$. 
Since
\begin{equation}
\label{ysca}
 y=a_{G_-}(\b,\g_-) + d_{G}(\g_-)\frac{\xi}{N^{\frac12}}
\equiv y_0+\d y, 
\end{equation}
the saddle point $p_c^G$ for $g_{y+1,\b}(z)$ is known to be 
\begin{equation}
  p_c^G\sim \g_-\left(1+\frac{\xi}{(\b_--\b)d_GN^{\frac12}}\right).
\end{equation}
Changing the path of $z$ in a way that it crosses $p_c^G$, 
\begin{equation}
  z=p_c^G\left(1+\frac{i w}{(\b_- - \b)d_GN^{\frac12}}\right)
=\g_-\left(1+\frac{iw+\xi}{(\b_- -\b)d_GN^{\frac12}}\right),
\end{equation}
we get
\begin{align}
&\quad g_{y+1,\b}(z)\notag\\
&=g_{y+1,\b}\big(p_c^G(y,\b)\big) 
+ \frac1{2!}g''_{y+1,\b}\big(p_c^G(y,\b)\big)
\left\{z-p_c^G(y,\b)\right\}^2
\notag\\
\label{gex}
&=g_{y_0+1,\b}\big(p_c^G(y_0,\b)\big)-\frac{d_G\xi\log\g_-}{N^{\frac12}}
-\frac{\xi^2+w^2}{2(\b_0-\b)N}. 
\end{align}
In the second equality we used
\begin{align}
 g_{y+1,\b}\big(p_c^G(y,\b)\big)
&\sim g_{y_0+1,\b}(p_c^G(y_0,\b))
+\left.\frac{\partial  g_{y+1,\b}}{\partial y}\right|_{y=y_0}\delta y
+\left.\frac1{2!}
\frac{\partial^2  g_{y+1,\b}}{\partial^2 y}\right|_{y=y_0}\delta y^2\notag\\
&=g_{y_0+1,\b}(p_c^G(y_0,\b))-\frac{d_G\xi\log\g_-}{N^{\frac12}}
-\frac{\xi^2}{2(\b_0-\b)N},
\end{align}
and
\begin{align}
  \frac1{2!}g''_{y+1,\b}\big(p_c^G(y,\b)\big)\left\{z-p_c^G(y,\b)\right\}^2
\sim\frac1{2!}g''_{y_0+1,\b}\big(p_c^G(y_0,\b)\big)\left\{z-p_c^G(y,\b)\right\}^2
=-\frac{w^2}{2(\b_0-\b)N}.
\end{align}
Combining the above, one finally finds
\begin{equation}
D_{1+}(u,x) \sim\mathcal{D}_{1+}(\b,\xi)\equiv
\frac{F(u,\g_-)}{(1-\g_+\g_-)\g_-^x}
\frac{e^{-\frac{\xi^2}{2(\b_--\b)}}}
{\sqrt{2\pi(\b_--\b)}d_G(\g_-)N^{\frac12}}.
\label{DD1+} 
\end{equation}
Similarly, one can show
\begin{align}
D_{1-}(u,x)&\sim\mathcal{D}_{1-}(\b,\xi)\equiv
\frac1{F(u,1/\g_+)(1-\g_+\g_-)\g_+^x}\frac{e^{-\frac{\xi^2}{2(\b-\b_+)}
\frac{d_G(\g_-)^2}{d_G(\g_+)^2}}}
{\sqrt{2\pi(\b-\b_+)}\frac{d_G(\g_+)}{d_G(\g_-)}d_G(\g_-)N^{\frac12}} 
\notag\\
&\hspace{54pt}\times e^{-\frac{a_{G-}(\g_-)-a_{G+}(\g_+)}{(\b-\b_+)}
\times\frac{d_G(\g_+)}{d_G(\g_-)}\xi N^{\frac12}} 
e^{-\frac{(a_G(\g_-)-a_G(\g_+))^2}{4(\b-\b_+)}
\frac{d_G(\g_+)}{d_G(\g_-)}\times N}.
\label{DD1-} 
\end{align}
From these relations the asymptotics of $\tilde{K}_{2N}$ is also obtained 
easily, 
\begin{equation}
\tilde{K}_{2N}(u_1,x_1;u_2,x_2)\sim (1-\g_+\g_-)\mathcal{D}_{1+}(\b_1,\xi_1)
\mathcal{D}_{1-}(\b_2,\xi_2).
\label{K2nasy}
\end{equation}
In fact one can find that~\eqref{K2nasy} does not contribute to
the Fredholm determinant by the following discussion. 
First we consider the asymptotics when $\b_+=\b_-=\b_0$ deforming the
contour of $z_i$ such that they cross the double critical point $p_c(\b_0)$,
\begin{align}
 \tilde{K}_{2N} (\b_0N,x_1;\b_0N,x_2)
& =
 \frac{1}{(2\pi i)^2}
 \int_{C_{R_1}} \frac{dz_1}{z_1} \int_{C_{R_2}} \frac{dz_2}{z_2}
 \frac{z_2^{x_2-N(\mu_c-1)}}{z_1^{x_1-N(\mu_c-1)}}
 e^{N(g_{\mu_c,\b_0}(z_1)+g_{\mu_c,\b_0}(1/z_2))} 
 \frac{z_2}{z_1-z_2},
\end{align}
where
\begin{align}
 z_1           
\sim p_c(\b_0)\left(1 - \frac{iw_1}{d(\b_0)N^{1/3}}\right), \quad
 \frac{1}{z_2} 
\sim  \frac{1}{p_c(\b_0)}\left(1 - \frac{iw_2}{d(\b_0)N^{1/3}}\right).
\end{align}
From~\eqref{Gb1}--~\eqref{z1z2asy2} and noticing
\begin{align}
& \frac{z_2^{x_2+N(1-\mu_c(\b_0))}}{z_1^{x_1+N(1-\mu_c(\b_0))}}
 \notag\\
&\sim p_c(\b_0)^{d_G N^{\frac12}(\xi_2-\xi_1)}
&\exp\left[iw_1(\frac{d_G}{d}N^{\frac16}\xi_1+\frac{a_{G-}-a}{d}N^{\frac23})
+iw_2(\frac{d_G}{d}N^{\frac16}\xi_1+\frac{a_{G-}-a}{d}N^{\frac23})\right],
\label{kdash}
\end{align}
we obtain
\begin{align}
 \tilde{K}_{2N}
 &\sim
 (p_c(\b_0))^{d_G N^{\frac12}(\xi_2-\xi_1)}
 \frac{1}{dN^{1/3}} \notag\\
&\quad
 \int_0^{\i} d\l 
\Ai(\frac{d_G}{d}N^{\frac16}\xi_1+\frac{a_{G-}-a}{d}N^{\frac23}+\l) 
\Ai(\frac{d_G}{d}N^{\frac16}\xi_2+\frac{a_{G-}-a}{d}N^{\frac23}+\l).
\label{kdashasy}
\end{align}
Using the asymptotics of Airy function
\begin{gather}
 \Ai(\frac{d_G}{d}N^{\frac16}\xi_1+\frac{a_{G-}-a}{d}N^{\frac23}+\l)
 \sim \frac1{2\sqrt{\pi}}
 \frac{e^{-\frac23N\frac{(a_{G-}-a)}{d}}}{\sqrt{\frac{a_{G-}-a}{d}
 N^{\frac23}}},
\label{AsAiry}
\end{gather}
where $a_{G-}-a>0$,
we can find $\tilde{K}_{2N}$ goes to $0$ asymptotically.
Combining this fact and the relation
\begin{align}
&\quad\tilde{K}_{2N}(\b_0N,x_1 ;\b_0N,x_2)\notag\\
&\sim (1-\g_+\g_-)\mathcal{D}_{1+}(\b_0,\xi_1)\mathcal{D}_{1-}(\b_0,\xi_2)
\notag\\
&\sim(1-\g_+\g_-)
\left[\frac1{(\g_+\g_-)^{a_{G-}(\g_-)}}\frac{(1-\a/\g_-)^{1+\b_0}}
{(1-\a\g_-)^{1-\b_0}}\frac{(1-\a/\g_+)^{1-\b_0}}{(1-\a\g_+)^{1+\b_0}}
\times e^{-\frac{(a_G(\g_-)-a_G(\g_+))^2}{4(\b_0-\b_+)}
\frac{d_G(\g_+)}{d_G(\g_-)}}\right]^{N},
\end{align}
one finds
\begin{equation}
 \left[\frac1{(\g_+\g_-)^{a_{G-}(\g_-)}}\frac{(1-\a/\g_-)^{1+\b_0}}
{(1-\a\g_-)^{1-\b_0}}\frac{(1-\a/\g_+)^{1-\b_0}}{(1-\a\g_+)^{1+\b_0}}
\times e^{-\frac{(a_G(\g_-)-a_G(\g_+))^2}{4(\b_0-\b_+)}
\frac{d_G(\g_+)}{d_G(\g_-)}}\right]<1.
\label{FRelation}
\end{equation}
Next we consider the case where $\b_1\neq\b_2$. From~\eqref{K2nasy},
it is straightforward to see
\begin{align}
  &\quad\tilde{K}_{2N}(\b_1N,x_1;\b_2N,x_2)\notag\\
&\sim (1-\g_+\g_-)\mathcal{D}_{1+}(\b_1,\xi_1)\mathcal{D}_{1-}(\b_2,\xi_2)
\notag\\
&\sim(1-\g_+\g_-)\times\g_-^{x_2-x_1}\big\{(1-\a/\g_-)(1-\a\g_-)
\big\}^{\b_1-\b_2}\notag\\ 
&\hspace{45pt}\times
\left[\frac1{(\g_+\g_-)^{a_{G-}(\g_-)}}\frac{(1-\a/\g_-)^{1+\b_2}}
{(1-\a\g_-)^{1-\b_2}}\frac{(1-\a/\g_+)^{1-\b_2}}{(1-\a\g_+)^{1+\b_2}}
\times e^{-\frac{(a_G(\g_-)-a_G(\g_+))^2}{4(\b_2-\b_+)}
\frac{d_G(\g_+)}{d_G(\g_-)}}\right]^{N}.
\label{FRelation2}
\end{align}
Hence from~\eqref{FRelation} and~\eqref{FRelation2} one finds that 
the contribution of $\tilde{K}_{2N}(\b_1N,x_1;\b_2N,x_2)$
is of order $\mathcal{O}(e^{-N})$ and is negligible in 
the Fredholm determinant of~\eqref{NewRep}. 
  
To obtain the asymptotics of ~\eqref{NewRep}, 
one also needs the asymptotics of $\phi$.
Let us represent $\phi_{r_1,r_2}(x_1,x_2)$ as 
\begin{equation}
\label{phi}
  \phi_{r_1,r_2}(x_1,x_2)=\frac{(1-\a)^{2(u_2-u_1)}}{2\pi i}\int_{C_1}
\frac{dz}{z}e^{Nf_{y_1,y_2}(z)},
\end{equation}
where 
\begin{equation}
  f_{y_1,y_2}(z)=g_{y_1+1,\b_1}(z)-g_{y_2+1,\b_2}(z).
\end{equation}
We scale $y_i$ as
\begin{equation}
  y_i=a_{G-}(\b_i,\g_-)+\frac{d_G(\g_-)\xi_i}{N^{\frac12}}\equiv y_{0}+\delta y_i,
\end{equation}
and adjust the path of $z$ such that it crosses the saddle point $p_c^f$ of 
$f_{y_1,y_2}(z)$,
\begin{equation}
  z=p_c^f\left(1+\frac{iw}{(\b_2-\b_1)d_GN^{\frac12}}\right)
\sim \g_-\left(1+\frac1{(\b_2-\b_1)d_GN^{\frac12}}(\xi_1-\xi_2+iw)\right).
\end{equation}
Then we get
\begin{equation}
\label{fex}
 f_{y_1,y_2}(z)\sim f_{y_0,y_0}(z_0)
 -\frac{d_G\log\g_-(\xi_1-\xi_2)}{N^{\frac12}}
 -\frac{(\xi_1-\xi_2)^2+w^2}{2(\b_2-\b_1)}.
\end{equation}
From~\eqref{phi} and~\eqref{fex}, one finds
\begin{equation}
 \phi_{r_1,r_2}(z) 
\sim\frac{\g_-^{x_2-x_1}}{d_GN^{\frac12}}
\left[\frac{(1-\a)^2}{(1-\a\g_-)(1-\a/\g_-)}\right]^{N(\b_2-\b_1)}
\frac{e^{\frac{-(\xi_2-\xi_1)^2}{2(\b_2-\b_1)}}}{\sqrt{2\pi(\b_2-\b_1)}}.
\label{phiasy}
\end{equation}

\medskip
Substituting these asymptotic forms to~\eqref{NewRep} and picking 
up the terms which do not vanish asymptotically, one finds
\begin{align}
&\quad\P[h^{+}(r_1,M)<l_1,\cdots ,h^{+}(r_m,M)<l_m]\notag\\
&\sim \det[1+ K_{2N}g]\left\{\sum_{k=1}^{m}(\g_+\g_-)^{l_k}
\frac{F(u_k,1/\g_+)}{F(u_k,\g_-)}\big(-1+\Lambda(\b,s)\big)\right\}\notag\\
&\quad+1-\sum_{k=1}^{m}\int_{s_k}^{\i}\frac{e^{-\frac{\xi_1^2}{2(\b_--\b_k)}}}
{\sqrt{2\pi(\b_--\b_k)}}d\xi_1 \notag\\
&\quad +
\sum_{i=1}^{m}\sum_{k_1<\cdots<k_i}
\int_{s_{k_1}}^{\i}\cdots
\int_{s_{k_i}}^{\i}d\xi_{1}\cdots d\xi_i
\frac{e^{\frac{-(\xi_2-\xi_1)^2}{2(\b_2-\b_1)}}}{\sqrt{2\pi(\b_2-\b_1)}}
\cdots
\frac{e^{\frac{-(\xi_i-\xi_{i-1})^2}{2(\b_{i}-\b_{i-1})}}}
{\sqrt{2\pi(\b_{i}-\b_{i-1})}}\frac{e^{-\frac{\xi_i^2}{2(\b_--\b_i)}}}
{\sqrt{2\pi(\b_--\b_i)}},
\label{Ph+asy}
\end{align}
where $(\g_+\g_-)^{l_k}
\frac{F(u_k,1/\g_+)}{F(u_k,\g_-)}\Lambda(\b,s)$ represents 
the contribution of the terms included in
$\sum D_-E'D_+$ such as
\begin{align}
&\sum_{k,k'}\sum_{x_1,x_2} D_{2-}(u_k,x_1)
\phi_{2u_k,2u_{k'}}(x_1,x_2)D_{2+}(u_{k'},x_2),\notag\\
&\sum_{k,k",k'}\sum_{x_1,x_2,x_3} D_{2-}(u_k,x_1)\phi_{2u_k,2u_{k"}}(x_1,x_2)
\tilde{K}_{2N}(u_{k"},x_3;u_{k'},x_2)D_{2+}(u_{k'},x_2),\notag\\
&\sum_{k,k",k'}\sum_{x_1,x_2,x_3}D_{2-}(u_k,x_1)
\phi_{2u_k,2u_{k"}}(x_1,x_3)\phi_{2u_{k"},
2u_{k'}}(x_3,x_2)D_{2+}(k',x_2), 
\end{align}
and so on. In fact the first term in~\eqref{Ph+asy} cancels due to 
the subtraction 
in~\eqref{ModDef};
\begin{align}
&\quad\det[1+ K_{2N}g]\notag\\
&\sim \det[1+  K_{2N}g]\left\{\sum_{k=1}^{m}(\g_+\g_-)^{l_k}
\frac{F(u_k,1/\g_+)}{F(u_k,\g_-)}\big(-1+\Lambda(\b,s)\big)\right\}\notag\\
&{\hspace{54pt}}-\g_+\g_-\det[1+ K_{2N}g_{l-1}]
\left\{\sum_{k=1}^{m}(\g_+\g_-)^{l_k-1}
\frac{F(u_k,1/\g_+)}{F(u_k,\g_-)}\big(-1+\Lambda(\b,s)\big)\right\}\notag\\
&\sim\frac{1}{d_G(\g_-)N^{\frac12}}\sum_{k=1}^{m} 
\left(\frac{\partial}{\partial s_k}\det[1+ K_{2N}g]\right)
\frac{F(u_k,1/\g_+)}{F(u_k,\g_-)}\big(-1+\Lambda(\b,s)\big)(\g_+\g_-)^{l_k}
\notag\\
&\sim\frac{1}{d_G(\g_-)N^{\frac12}}\sum_{k=1}^{m} 
\tilde{K}_{2N}
(u_k,x=a_{G-}N+d_GN^{\frac12}s_k;u_k,x=a_{G-}N+d_GN^{\frac12}s_k)
\frac{F(u_k,1/\g_+)}{F(u_k,\g_-)}\notag\\
&\quad\big(-1+\Lambda(\b,s)\big)(\g_+\g_-)^{l_k}
\notag\\
&\sim 0,
\end{align}
where $g_{l-1}(r_j,x_j)=-\chi_{(l_j-1,\i)(x_j)}$.
Thus when we consider the asymptotic behavior of~\eqref{ModDef}, 
$\P[h^+(r_1,M),\cdots ,h^{+}(r_m,M)]$ can be replaced with 
\begin{align}
&\quad\P'[h^+(r_1,M),\cdots ,h^{+}(r_m,M)]\notag\\
&=
1-\sum_{k=1}^{m}\int_{s_k}^{\i}d\xi_k\frac{e^{-\frac{\xi_k^2}{2(\b_--\b_k)}}}
{\sqrt{2\pi(\b_--\b_k)}}\notag\\
&\quad +
\sum_{i=1}^{m}\sum_{k_1<\cdots<k_i}
\int_{s_{k_1}}^{\i}\cdots
\int_{s_{k_i}}^{\i}d\xi_{1}\cdots d\xi_i
\frac{e^{\frac{-(\xi_2-\xi_1)^2}{2(\b_2-\b_1)}}}{\sqrt{2\pi(\b_2-\b_1)}}
\cdots
\frac{e^{\frac{-(\xi_i-\xi_{i-1})^2}{2(\b_{i}-\b_{i-1})}}}
{\sqrt{2\pi(\b_{i}-\b_{i-1})}}\frac{e^{-\frac{\xi_i^2}{2(\b_--\b_i)}}}
{\sqrt{2\pi(\b_--\b_i)}}.
\end{align} 
Using the fact that~\eqref{ModDef} can be expressed as
\begin{align}
&\quad\P[h(r_1,M)<l_1,\cdots ,h(r_m,M)<l_m] \notag\\
&\sim\left(1+\frac1{(1-\g_+\g_-)d_GN^{\frac12}}\sum_{k=1}^{m}
\frac{\partial}{\partial s_k}\right)
\P'[h^+(r_1,M)<l_1,\cdots ,h^+(r_m,M)<l_m]\notag\\
&\sim \P'[h^+(r_1,M)<l_1,\cdots ,h^+(r_m,M)<l_m],
\end{align}
one finally gets
\begin{align}
&\quad\P[h(r_1,M)<l_1,\cdots ,h(r_m,M)<l_m]\notag\\
&\sim1-\sum_{k=1}^{m}\int_{s_k}^{\i}d\xi_k\frac{e^{-\frac{\xi_k^2}{2(\b_--\b_k)}}}
{\sqrt{2\pi(\b_--\b_k)}}\notag\\
&\quad +
\sum_{i=1}^{m}\sum_{k_1<\cdots<k_i}
\int_{s_{k_1}}^{\i}\cdots
\int_{s_{k_i}}^{\i}d\xi_{1}\cdots d\xi_i
\frac{e^{\frac{-(\xi_2-\xi_1)^2}{2(\b_2-\b_1)}}}{\sqrt{2\pi(\b_2-\b_1)}}
\cdots
\frac{e^{\frac{-(\xi_i-\xi_{i-1})^2}{2(\b_{i}-\b_{i-1})}}}
{\sqrt{2\pi(\b_{i}-\b_{i-1})}}\frac{e^{-\frac{\xi_i^2}{2(\b_--\b_i)}}}
{\sqrt{2\pi(\b_--\b_i)}}.
\end{align}
This expression is the same as the Fredholm representation 
in~\eqref{gausuasy} with the kernel (\ref{kernelG}).
\qed
\vspace{18pt}

We end this chapter by providing a proof of (ii-a) in Theorem 2.1,
for which the same strategy as the above proof of (i-c),(ii-a) and 
(iii-a) is applicable.
The only difference is the asymptotics of $D_{1-}$
\eqref{DD1-}. This changes to
\begin{align}
  D_{1-}(u_c=\b_c N,x)&\sim\mathcal{D'}_{1-}(\b_c,\xi)\equiv
\frac1{F(u_c,1/\g_+)(1-\g_+\g_-)\g_+^x}\frac{e^{-\frac{\xi^2}{2(\b_c-\b_+)}
\frac{d_G(\g_-)^2}{d_G(\g_+)^2}}}
{\sqrt{2\pi(\b-\b_+)}\frac{d_G(\g_+)}{d_G(\g_-)}d_G(\g_-)N^{\frac12}} 
.
\end{align}
This means that we can obtain the Gaussian as a scaling limit 
for both $D_{1+}(u,x)$
and $D_{1-}(u,x)$ since the $\b_c$ point is the crossing point of two
lines with Gaussian fluctuation. Together with this, the asymptotics of 
$\tilde{K}_{2N}$ also change;
\begin{align}
\tilde{K}_{2N}(u_c,x_1;u_c,x_2)\sim (1-\g_+\g_-)\mathcal{D}_{1+}(\b_c,\xi_1)
\mathcal{D}'_{1-}(\b_c,\xi_2)\sim \mathcal{O}(e^{-N}).  
\end{align}
Considering these two relations in addition to~\eqref{DD1+} and~\eqref{phiasy}, we pick up the terms which do not vanish 
in~\eqref{NewRep},
\begin{align}
&\quad\P[h^{+}(r_c=2u_c,M)<l]\notag\\
&\sim \det[1+g K_{2N}]\left\{(\g_+\g_-)^{l}
\frac{F(u_c,1/\g_+)}{F(u_c,\g_-)}\left(-1+\Lambda(\b_c,s)\right)\right\}\notag\\
&\quad+1-\int_{s}^{\i}d\xi_1\frac{e^{-\frac{\xi_1^2}{2(\b_--\b_c)}}}
{\sqrt{2\pi(\b_--\b_c)}}-\int_{s}^{\i}d\xi_2
\frac{e^{-\frac{\xi_2^2}{2(\b_c-\b_+)}
\frac{d_G(\g_-)^2}{d_G(\g_+)^2}}}
{\sqrt{2\pi(\b_c-\b_+)}\frac{d_G(\g_+)}{d_G(\g_-)}} 
\notag\\
&\quad +\int_{s}^{\i}d\xi_1\frac{e^{-\frac{\xi_1^2}{2(\b_--\b_c)}}}
{\sqrt{2\pi(\b_--\b_c)}}\int_{s}^{\i}d\xi_2
\frac{e^{-\frac{\xi_2^2}{2(\b_c-\b_+)}
\frac{d_G(\g_-)^2}{d_G(\g_+)^2}}}
{\sqrt{2\pi(\b_c-\b_+)}\frac{d_G(\g_+)}{d_G(\g_-)}}
 ,
\label{PPh+asy}
\end{align}
Here we can neglect the first term 
since this term vanishes due to the subtraction
in~\eqref{ModDef}. Hence one obtains, 
\begin{align}
 \lim_{N\to\i} \P[H_N^{(G_-)}(\b_c,\g_{-})\leq s]
 =\int_{-\i}^{s}d\xi_1\frac{e^{-\frac{\xi_1^2}{2(\b_--\b_c)}}}
{\sqrt{2\pi(\b_--\b_c)}}\int_{-\i}^{s}d\xi_2
\frac{e^{-\frac{\xi_2^2}{2(\b_c-\b_+)}
\frac{d_G(\g_-)^2}{d_G(\g_+)^2}}}
{\sqrt{2\pi(\b_c-\b_+)}\frac{d_G(\g_+)}{d_G(\g_-)}}.
\end{align}
This completes the proof of (ii-a) in Theorem 2.1. 
In principle, one can also apply the same method to the case of 
the multipoint function
including the $\b_c$ point.

\setcounter{equation}{0}
\section{Transition around GOE$^2$}
\label{goe2}
\subsection{Limiting Kernel}
Let us suppose that, when $\g_-=\g_0$, the limit shapes of $a(\b)$ 
and $a_{G-}(\b, \g_-)$
cross at $\b=\b_0$. We call this point the GOE$^2$ point.

In this section, we obtain the kernel describing the 
multi-point height fluctuation near this GOE$^2$ point.
We consider the case where the parameter $\a \leq \g_+<1/\g_0$ 
is fixed and $\g_-$ scales like
\begin{equation}
\label{g-scale}
 \g_- = \g_0\left( 1-\frac{\o}{d(\b_0)N^{1/3}}\right),
\end{equation} 
with $\o$ fixed.
The result is 

\begin{theorem}
\begin{equation}
\label{detKlim}
 \lim_{N\to\i} \P[H_N(\t_1,\b_0) \leq s_1, \cdots , H_N(\t_m,\b_0) \leq s_m]
 =
 \det(1+\K \G),
\end{equation}
where $\G(\t_j,\xi)=-\chi_{(s_j,\i)}(\xi)$ ($j=1,2,\cdots,m$) and 
\begin{align}
 &\quad 
 \K(\t_1,\xi_1;\t_2,\xi_2) \notag\\
 &=
 \begin{cases}
  \K_2(\t_1,\xi_1;\t_2,\xi_2)
  +\Ai(\xi_1) \int_0^{\i} d\l e^{-(\o+\t_2)\l}\Ai(\xi_2-\l), 
  & \o+\t_2>0 ,   \\
  \K_2(\t_1,\xi_1;\t_2,\xi_2)
  -\Ai(\xi_1) \int_0^{\i} d\l e^{(\o+\t_2)\l}\Ai(\xi_2+\l)   
  +\Ai(\xi_1) e^{\frac{\t_2^3+\o^3}{3}-\xi_2(\t_2+\o)},
  & \o+\t_2<0 .
 \end{cases}
\label{Kernel12}
\end{align}
\end{theorem}
\noindent

\vspace{3mm}\noindent
{\bf Remark.} 
This kernel seems to be new.
If we set $\o=0$ from the beginning, this gives 
Theorem 3.1 (i-b). 
Notice that when we focus on the one-point correlation function,
the case where $\t =0, \o=\t'$ is essentially 
the same as the case where $\t=\t' ,\o=0$.


\medskip
\noindent
{\bf Proof.} 
First we give the contour integral representation of the kernel.
For $\g_-<R_2$, it reads
\begin{align}
 &\quad 
 \tilde{K}_N(r_1=2u_1,x_1;r_2=2u_2,x_2) \notag\\
 &=
 \frac{ (1-\a)^{2(u_2-u_1)}}{(2\pi i)^2}
 \int_{C_{R_1}} \frac{dz_1}{z_1}\int_{C_{R_2}}\frac{dz_2}{z_2}
 \frac{z_2^{x_2}}{z_1^{x_1}} \frac{z_1}{z_1-z_2}
 \frac{(1-\a/z_1)^{N-1+u_1}(1-\a z_2)^{N-1-u_2}}
      {(1-\a z_1)^{N-1-u_1}(1-\a/z_2)^{N-1+u_2}}
 \notag\\
 &\quad \times
 \frac{1-\g_-/z_1}{1-\g_-/z_2} \frac{1-\g_+ z_2}{1-\g_+ z_1}.
\label{Kn1}
\end{align}
For $\g_->R_2$, we have to add the contribution of the pole 
at $z_2=\g_-$, resulting in
\begin{align}
 &\quad 
 \tilde{K}_N(r_1=2u_1,x_1;r_2=2u_2,x_2) \notag\\
 &=
 \frac{ (1-\a)^{2(u_2-u_1)}}{(2\pi i)^2}
 \int_{C_{R_1}} \frac{dz_1}{z_1}\int_{C_{R_2}}\frac{dz_2}{z_2}
 \frac{z_2^{x_2}}{z_1^{x_1}} \frac{z_1}{z_1-z_2}
 \frac{(1-\a/z_1)^{N-1+u_1}(1-\a z_2)^{N-1-u_2}}
      {(1-\a z_1)^{N-1-u_1}(1-\a/z_2)^{N-1+u_2}}
 \notag\\
 &\quad \times
 \frac{1-\g_-/z_1}{1-\g_-/z_2}  \frac{1-\g_+ z_2}{1-\g_+ z_1}
 \notag\\
 &\quad 
 +
 \frac{ (1-\a)^{2(u_2-u_1)}}{2\pi i}
 \int_{C_{R_1}} \frac{dz_1}{z_1} \frac{1}{z_1^{x_1}}
 \frac{(1-\a/z_1)^{N-1+u_1} \g_-^{x_2}}{(1-\a z_1)^{N-1-u_1}(1-\g_+z_1)}   
 \frac{(1-\a\g_-)^{N-1-u_2}(1-\g_+\g_-)}{(1-\a/\g_-)^{N-1+u_2}}.
\label{Kn2}
\end{align}

Now we consider the asymptotics when we set (\ref{g-scale}),
\begin{align}
 x_i &= N a(\b_0) + d(\b_0) N^{1/3} \xi_i, \\
 r_i &= 2 u_i = 2N\left( \b_0+\frac{c(\b_0)\t_i}{N^{1/3}} \right),
\end{align}
with $i=1,2$ and take $N\to\i$.
The analysis is almost the same as in \cite{SI2003p}
and hence the details are omitted.
When $\o+\t_2>0$, one uses (\ref{Kn1}) and 
deforms the contours of $z_1$ and $z_2$ such that 
\begin{align}
 \label{pathz12}
 z_1          
&\sim \g_0\left(1 + \frac{1}{d(\b_0)N^{1/3}}(\t_1-iw_1)\right), \\
 \label{pathz22}
 \frac{1}{z_2}
&\sim  \frac1{\g_0}\left(1 - \frac{1}{d(\b_0)N^{1/3}}(\t_2+iw_2)\right).
\end{align}
Then we get
\begin{align}
 K_N 
 &\sim
 (1-\a)^{2(u_2-u_1)} (p_c(\b_0))^{(\xi_2-\xi_1)d(\b_0)N^{1/3}}
 \frac{1}{d(\b_0)N^{1/3}} \notag\\
 &\quad\times
 e^{ \l_1(\b_0)c(\b_0)N^{2/3}(\t_1-\t_2)
    +\l_2(\b_0)c(\b_0)^2 N^{1/3}(\t_1^2-\t_2^2)
    +\l_3(\b_0)c(\b_0)^3 (\t_1^3-\t_2^3)+\xi_2 \t_2-\xi_1 \t_1} \notag\\
 &\quad\times
 \frac{1}{4\pi^2}
 \int_{\text{Im}w_1=\eta_1} dw_1 \int_{\text{Im}w_2=\eta_2} dw_2 
 \left(-\frac{1}{\tau_2-\tau_1+i(w_1+w_2)}+\frac{1}{\o+\tau_2+iw_2}\right)
 \notag\\
 &\quad\times
 e^{i\xi_1 w_1 + i\xi_2 w_2 + \frac{i}{3}(w_1^3+w_2^3)}, 
\end{align}
with the definition of $\l_i$ given in~\eqref{deflam}.
Noticing 
\begin{align}
 &\quad 
 \frac{1}{4\pi^2}
 \int_{\text{Im}w_1=\eta_1} dw_1 \int_{\text{Im}w_2=\eta_2} dw_2 
 \left(-\frac{1}{\tau_2-\tau_1+i(w_1+w_2)}+\frac{1}{\o+\tau_2+iw_2}\right) 
 \notag\\&\quad\times
 e^{i\xi_1 w_1 + i\xi_2 w_2 + \frac{i}{3}(w_1^3+w_2^3)} 
 \notag\\
 &=
 \int_0^{\i} e^{-\l(\t_1-\t_2)}\Ai(\xi_1+\l)\Ai(\xi_2+\l)d\l
 +\Ai(\xi_1) \int_0^{\i} d\l e^{-(\o+\t_2)\l}\Ai(\xi_2-\l) ,
\end{align}
one gets the desired expression for this case.

The $\o+\t_2<0$ case can also be treated in a similar fashion.
In this case we consider the asymptotics using
~\eqref{Kn2} since the contour of $z_2$~\eqref{pathz22} 
cross the real axis on the left of $\g_0$.
The second term in the RHS of~\eqref{Kn2} produces the third term 
in the RHS of~\eqref{Kernel12}.
\qed

\subsection{Connection to the Baik-Rains analysis}

When we specialize to the case of the one-point height fluctuation, 
our formula reduces to 
\begin{equation}
\label{Ho}
 \lim_{N\to\i} \P[H_N(0,\b_0) \leq s]
 =
 \det(1+\K\G),
\end{equation}
where  $~~\G(\xi)=-\chi_{(s,\i)}(\xi)$ and 
\begin{equation}
 \K(x,y) = \K_2(x,y) + A(x) B(y,\o),
\end{equation}
with 
\begin{align}
& \K_2(x,y) = \int_0^{\i} \Ai(x+\l)\Ai(y+\l)d\l,\hspace{37mm} \\
& A(x)      = \Ai(x),
\end{align}
 \begin{subnumcases}{\label{defB}B(x,\o)=}
               \int_0^{\i} d\l e^{-\o\l}\Ai(x-\l) , & $\o>0$ \label{416a}\\
               -\int_0^{\i} d\l e^{\o\l} \Ai(x+\l)+e^{\o^3/3-x\o},  & 
       $\o<0 .$ \label{416b}
 \end{subnumcases}
Notice that the expression on the right hand side of 
(\ref{Ho}) is independent of $\b_0$ and that, 
for the special case where $\b_0=0$, 
another expression for the same quantity was previously obtained
in \cite{BR2000}. Hence the two expressions should be the same.
For $\o=0$, this was already shown in \cite{Forrester2000p}.
In this subsection, we prove the equivalence for any $\o$ 
generalizing the arguments in \cite{Forrester2000p}. 

We first proceed as 
\begin{align}
 \det(1+\K\G) 
 &= 
 \det(1+(\K_2+A\otimes B)\G) \notag\\
 &=
 \det(1+\K_2\G) \det(1+(1+\K_2\G)^{-1}A\otimes B\G) \notag\\
 &=
 F_2(s)\left(1-\int_s^{\i}dx\int_s^{\i}dy \r(x,y)A(x) B(y,\o)\right),
\end{align}
where $\r(x,y)$ is the kernel of the operator $(1+\K_2\G)^{-1}$.
Hence if one defines 
\begin{equation}
\label{aFred}
 a(s,\o) = 1-\int_s^{\i}dx\int_s^{\i}dy \r(x,y)A(x) B(y,\o),
\end{equation}
one has
\begin{equation}
 \det(1+\K\G) = F_2(s) a(s,\o)\equiv F(s).
\label{F(s)}
\end{equation}
Let us also define 
\begin{equation}
\label{bFred}
 b(s,\o) = \int_s^{\i} dy \r(s,y) B(y,\o),
\end{equation}
and
\begin{align}
 Q(x) &= \int_s^{\i} dy \r(x,y) A(y), \\
 q(s) &= Q(s).
\end{align}
Now we show that the functions $a,b$ have the following
properties.
\begin{proposition}
\begin{align}
\frac{\partial}{\partial s}a  &=  q b, \label{a_s} \\
\frac{\partial}{\partial s}b  &=  q a - \o b, \label{b_s} \\
\frac{\partial}{\partial \o}a &=  q^2 a  -(q'+\o q)b, \label{a_o} \\
\frac{\partial}{\partial \o}b &=  (q'-\o q) a +(\o^2-s-q^2) b \label{b_o}.
\end{align}
\end{proposition}

\medskip
\noindent
{\bf Proof.}
From the results in \cite{TW1994}, one has
\begin{gather}
 R(x,y)\equiv \r(x,y)-\d^+(x-y)=\frac{Q(x)P(y)-P(x)Q(y)}{x-y},
\label{Rdef}\\
 \pd{s} Q(y) 
 = 
 q(s) \big(\d^+(y-s)-\r(s,y)\big), \\
 \left(\pd{s}+\pd{x}+\pd{y}\right)\r(x,y) = -Q(x) Q(y),\\
 Q'(y) = P(y) -u(s) Q(y) + q(s) \r(s,y) - q(s)\d^+(y-s),
\label{TW3} \\
 P'(y) = y Q(y)-2 Q(y) v(s)+ u(s) P(y) +  p(s) \r(s,y) - p(s)\d^+(y-s) ,
\label{TW4}\\
u^2-2v=q^2,
\label{TW5}\\
q'=p-qu,
\label{TW6}
\end{gather}
where
\begin{align}
 &P(x) = \int_s^{\i} \r(x,y) \Ai'(y) dy, \\ 
 &p(s) = P(s),\\
 &u(s) = \int_s^{\i} Q(x) \Ai(x) dx, 
  \label{u(s)}\\
 &v(s) = \int_s^{\i} P(x) \Ai(x) dx, \\
 &\int_s^{\i} \d^+(y-s) f(y) dy = f(s).
\end{align}
Besides, we also use 
\begin{gather}
 \frac{\partial}{\partial y}B(y,\o) = \Ai(y) -\o B(y,\o), 
\label{SI1}\\
 \pd{\o}B(y,\o)
 =
 \Ai'(y) -\o \Ai(y) + (\o^2-y) B(y,\o).
\label{SI2}
\end{gather}
These equations can be shown immediately.
For the case where $\o>0$ we use \eqref{416a} and compute
\begin{align}
  &\frac{\partial}{\partial y}B(y,\o)=\int_0^{\infty}d\l
e^{-\o\l}\frac{\partial \Ai(y-\l)}{\partial y} =
-\int_0^{\infty}d\l
e^{-\o\l}\frac{\partial \Ai(y-\l)}{\partial \l}, \\
  &\frac{\partial}{\partial\o}B(y,\o)=-\int_0^{\infty}
d\l e^{\o\l}\l \Ai(y-\l)\notag\\
&~~~~~~~~~~~~~~~~
=-\int_0^{\infty}d\l e^{-\o\l}\big\{y \Ai(y-\l)- \Ai''(y-\l)\big\}. 
\label{Bo} 
\end{align}
In~\eqref{Bo} the Airy equation, $\Ai''(x)=x\Ai(x)$, is used. 
These equations lead to~\eqref{SI1} and~\eqref{SI2} respectively.
For the case where $\o<0$ they can be 
shown by applying the same method to \eqref{416b}. 

Now the first two equalities, (\ref{a_s}) and (\ref{b_s}), 
can be shown as
\begin{align}
 \pd{s}a 
 &=
 Q(s) B(s,\o) - \int_s^{\i} dy \pd{s} Q(y) B(y,\o) \notag\\
 &=
 q(s) \int_s^{\i} dy \r(s,y) B(y,\o) 
 =
 q b, \\
 \pd{s}b
 &=
 -\int_s^{\i} dy \left(\pd{y} \r(s,y)\right) B(y,\o) 
 -q(s) \int_s^{\i} dy Q(y) B(y,\o) -\r(s,s)B(s,\o)\notag\\
 &=
 q a -q  + \int_s^{\i} dy \r(s,y) \pd{y} B(y,\o) \notag\\
 &=
 qa -\o b. 
\end{align}
For the third equality (\ref{a_o}), one starts from 
\begin{align}
 \pd{\o}a
 &=
 -\int_s^{\i} Q(y) \pd{\o} B(y,\o) dy \notag\\
 &=
 -\int_s^{\i} Q(y) \Ai'(y) dy + \o \int_s^{\i} Q(y) \Ai(y) dy
 \notag\\
 &\quad 
 -\o^2\int_s^{\i} Q(y) B(y,\o) dy + \int_s^{\i} y Q(y) B(y,\o)dy.
\end{align}
The second and the third terms are easily seen to be
$\o u$ and $-\o^2(1-a)$ respectively. Using~\eqref{TW3},
the first term is computed as
\begin{align}
 -\int_s^{\i} Q(y) \Ai'(y) dy
 &=
 -\big[Q(y) \Ai(y)\big]_s^{\i}
 +
 \int_s^{\i} Q'(y) \Ai(y) dy \notag\\
 &=
 \int_s^{\i} P(y) \Ai(y) dy -u \int_s^{\i} Q(y) \Ai(y) dy
 +
 q \int_s^{\i}\r(s,y) \Ai(y) dy \notag\\
 &=
 v-u^2+q^2.
\end{align}

Next we consider the last term. First we calculate
\begin{align}
yQ(y)&=P'(y)-uQ'(y)-Q(y)(u^2-2v)+(\r-\d^+)(qu-p)\notag\\
     &=P'(y)-uQ'(y)-Q(y)q^2 - \r q'+\d^+q',
\label{yQ}
\end{align}
where we use~\eqref{TW3},~\eqref{TW4} in the first equality and 
~\eqref{TW5},~\eqref{TW6} in the second equality. Then 
one can show that the last term is 
\begin{align}
 \int_s^{\i} dy y Q(y) B(y,\o)
 &=-q^2\int_s^{\infty}dyQ(y)B(y,w)-q'b+\int_s^{\infty}dyP'(y)B(y)\\
&\quad -u\int_s^{\infty}dyQ'(y)B(y,w)+q'B(s,w)\notag\\
 &=q^2 a-(q'+\o q)b -v+u^2-q^2-\o u+\o^2(1-a),
\end{align}
where we use \eqref{SI1} and \eqref{SI2}.
Combining these, one gets (\ref{a_o}).

Finally, for the proof of the last equality (\ref{b_o}), one 
starts from 
\begin{align}
 \pd{\o}b
 &=
 \int_s^{\i} \r(s,y) \pd{\o} B(y,\o) dy \notag\\  
 &=
 \int_s^{\i} dy \r(s,y) \Ai'(y) 
 -
 \o\int_s^{\i} dy \r(s,y) \Ai(y)
 \notag\\
 &\quad
 +\o^2 \int_s^{\i} dy \r(s,y) B(s,\o)
 -
 \int_s^{\i} dy y \r(s,y) B(s,y).
\end{align}
The first three terms are easily seen to be
$p=q'+u q, -\o q, \o^2 b$ respectively.
The last term can also be calculated. The term $y\r$ in the integrand
can be shown as
\begin{gather}
  y\r(s,y)=(p-qu)Q(y)-qQ'(y)+(q^2+s)\r(s,y)-q^2\d^+(s-y)
\label{yr}
\end{gather}
due to~\eqref{Rdef} and \eqref{TW3}. Thus from~\eqref{TW6},~\eqref{SI1} and 
~\eqref{yr}, one gets
\begin{gather}
  -\int_s^{\infty}dy y\r(s,y)B(s,y)=-(q'-\o q)\int_{s}^{\infty}dy Q(y)B(y)-
(q^2+s)b-uq
\end{gather}
These equations lead to (\ref{b_o}).
\qed

This proposition, combined with 
$a(s,0)=b(s,0)=e^{-\int_s^{\i}q(x)dx}$ proved in \cite{Forrester2000p},
shows that the functions $a,b$ as defined by 
(\ref{aFred}),(\ref{bFred}) are the same as those in \cite{BR2000}
with the identification $a_{\text{BR}}=a,b_{\text{BR}}=-b,u=-q,w=2\o$.
Thus the one-point height fluctuation defined in \eqref{F(s)} corresponds
to that in \cite{BR2000} under the above identification.

\subsection{GOE$^2$ to GUE/Gaussian Transition}

Using the above results, it is possible to study the 
fluctuation properties of the PNG model quite in detail.
In this subsection, we set $\o=0$ and consider the transition
from GOE$^2$ ($\t=0$) to GUE ($\t\to\i$) and Gaussian ($\t\to -\i$).
As an example, we here consider the average of the scaled height.
For the case where $\t=0$ or $\t=\i$,
we can easily compute the average numerically using the 
Painlev\'e expression of the GOE$^2$ or GUE \cite{TW1994,TW1996}.  
For the other values of $\t$, however, such a representation
has not been known. 
To compute the value numerically, one can use the differential 
equations of $a$ and $b$,~\eqref{a_s}--\eqref{b_o}.
On the other hand, we also obtain the asymptotic behaviors of 
the average for each case where $\t\sim 0, \infty, -\infty$ as follows.
\begin{flushleft}
 {\bf{$\bullet~~{\t\sim 0}$}}
\end{flushleft}
The equations~\eqref{a_s}--\eqref{b_o} enable us to know the behaviors of 
$a$ for $\t\sim 0$. We get
\begin{equation}
\label{at}
 a(s,\t)
 =
 a(s,0)+\left.\frac{\partial a(s,\t)}{\partial\t}\right|_{\t=0}\t
 +\frac12\left.\frac{\partial^2a(s,0)}{\partial\t^2}\right|_{\t=0}
 \t^2
 +\frac1{3!}
 \left.\frac{\partial^3a(s,\t)}{\partial\t^3}\right|_{\t=0}\t^3 +\cdots .
\end{equation}
hen the distribution behaves as
%
Substituting~\eqref{at}  
to $F(s)$ defined in \eqref{F(s)}, the average behaves as
\begin{align}
 \int s F'(s) ds
 &\sim
 \int s F_{\text{GOE}^2}'(s)ds 
 +\t\int e_1(s)ds + \frac{\t^2}{2!}\int e_2(s)ds
 +\frac{\t^3}{3!}\int e_3(s)ds +\cdots\notag\\
 &= -0.49364-0.89941\t+0.41582\t^2-0.12409\t^3 +\cdots.
\label{average1}
\end{align}
Here $e_i(s)$'s are computed as
\begin{align}
e_1(s)=se^{-\int_s^{\infty}q(x)dx}&\left\{P_2(s)(q(s)^2-q(s)')-F_2(s)
\left( q(s)^3 + sq(s) - q'(s)q(s) \right)  \right\},
\\
e_2(s)=se^{-\int_s^{\infty}q(x)dx}&\left\{
P_2(s)\left(q(s)\,\left( -q(s) + q(s)^4 + 2\,q(s)^2\,s + s^2 - s\,q'(s) - 
    {q'(s)}^2 \right)\right)\right. \notag\\
&\left.
+F_2(s)\left(-q(s) + q(s)^4 + s\,q'(s) - {q'(s)}^2\right)
\right\},
\\
e_3(s)=se^{-\int_s^{\infty}q(x)dx}&\left\{
P_2(s)\left(q(s)^3 + q(s)^6 + 2\,q(s)\,s - 
  \left( q(s) + q(s)^4 + q(s)^2\,s + s^2 \right) \,q'(s) \right.\right.
\notag\\
&+ \left.
  \left( -q(s)^2 + s \right) \,{q'(s)}^2 + {q'(s)}^3\right)\notag\\
&+F_2(s)\left(
q(s)\,\left( 2 - q(s)^3 - q(s)^6 + q(s)\,s - 3\,q(s)^4\,s - 
    3\,q(s)^2\,s^2 - s^3 + \right.\right.\notag\\
&\left.\left.\left.    \left( q(s) + q(s)^4 + q(s)^2\,s + s^2 \right) \,q'(s)
+ 
    \left( q(s)^2 + 2\,s \right) \,{q'(s)}^2 - {q'(s)}^3
    \right)
\right)
\right\},
\end{align}
using~\eqref{a_s}--~\eqref{b_o} and 
$a(s,0)=b(s,0)=e^{-\int_s^{\i}q(x)dx}$ proved in\cite{Forrester2000p}.

\begin{flushleft}  
{\bf{$\bullet~~{\t\rightarrow \i}$}}
\end{flushleft}
One can also get the asymptotic behavior of $a$ as $\t\to\i$.
By noting
\begin{align}
 B(x,\t) &= 
 \frac{1}{\t}\int_0^{\i}d\th e^{-\th} \Ai(x-\th/\t) \notag\\ 
 &=
 \sum_{j=0}^{\i} \frac{(-1)\Ai^{(j)}(x)}{\t^{j+1}} \int_0^{\i}d\th \th^j e^{-\th}
 \notag\\
 &\sim 
 \frac{\Ai(x)}{\t} -\frac{\Ai'(x)}{\t^2}+\frac{\Ai''(x)}{\t^3}-\cdots ,
\end{align}
one finds
\begin{align}
 a(x,\t) 
 &\sim
 1-\frac{1}{\t}\int_s^{\i}dx\int_s^{\i}dy \r(x,y) \Ai(x) \Ai(y)
 = 
 1-\frac{u(s)}{\t},
\label{a/t}
\end{align}
where $u(s)$ defined in \eqref{u(s)} reads 
\begin{equation}
 u(s) = \int_s^{\i} q(x)^2 dx.
\end{equation}
Then one has
\begin{equation}
\int s F'(s) ds
 \sim \int s F_2'(s)ds  -\frac{1}{\t} \int F_2(s) u(s)ds
\sim -1.77109 + \frac{1}{\t}.
\label{average2}
\end{equation}

\begin{flushleft}
{\bf{$\bullet~~{\t\rightarrow -\i}$}}
\end{flushleft}
Finally we consider the asymptotics where $\t\sim -\infty$.
Due to~\eqref{a_s}, the probability density is represented as
\begin{equation}
\frac{d F(s)}{d s}
=q(s) b(s,\t)F_2(s)+ a(s,\t)F'_2(s). 
\label{PD} 
\end{equation}
We evaluate the first term of the above equation as follows. 
From the results of Baik-Rains,
\begin{align}
  &\lim_{\t\rightarrow\infty} a(s,\t)=1, \notag\\
  &\lim_{\t\rightarrow -\infty}b(s,\t)=0,\notag\\
  &a(s,\t)=b(s,-\t)e^{\t^3-s\t},
\end{align}
one finds when $\t\rightarrow\infty$,
\begin{equation}
  b(s,\t)\sim e^{\t^3-s\t}.
 \label{basymptotics}
\end{equation}
Let us scale the variable $s$ as
\begin{equation}
  s=\t^2 y,
\end{equation}
and we assume $y>0$. (This can be justified in the below discussion.)
Using the asymptotic behavior of the Airy function,
we also find
\begin{equation}
  q(s)\sim \Ai(s)\sim \frac{1}{2\sqrt{\pi}s^{1/4}}{e^{-\frac23s^{\frac32}}}.
\label{qasymptotics}
\end{equation}
Then due to~\eqref{basymptotics},~\eqref{qasymptotics} and 
$F_2(s)\sim 1$ as $s\rightarrow\infty$, one finds
\begin{equation}
 q(s)b(s,\t)F_2(s)
\sim \frac{1}{2\sqrt{\pi}|\t|^{1/2}y^{1/4}}e^{\t^3 g(y)},
\end{equation}
where
\begin{equation}
 g(y) = \frac{2}{3}y^{3/2}-y+1.
\end{equation}
The condition $g'(y)=0$ leads to $y=1$. 
Rescaling $y$ as $y=1+y'/|\t|^{3/2}$ and expanding $g(y)$ around $y=1$, 
one gets
\begin{equation}
 q(s)b(s,\t)F_2(s) 
 \sim
 \frac{1}{2\sqrt{\pi}|\t|^{1/2}} e^{\frac{y'^2}{4}}.
\end{equation}
On the other hand, for the second term, one finds
\begin{equation}
 a(s,\t)F'_2(s)  
 =
 a(\t^2 y, \t) F'_2(\t^2 y)
 \sim 
 0
\end{equation}
since $a(s,\t)\sim 1, F'_2(s)\to 0$ as $s\to\infty$.
Thus the probability density is approximated by the Gaussian
with the peak at $s=\t^2$ when $-\t$ is large sufficiently and 
one gets
\begin{equation}
  \int sF_2'(s) ds \sim \t^2.
\label{average3}
\end{equation}


\vspace{9pt}
Using~\eqref{average1} ,~\eqref{average2} and~\eqref{average3},
we can calculate the average
of the distribution numerically. 
This is shown in Fig 3.
In principle, other statistical quantities such as 
two point function analyzed in \cite{PS2003} can be also studied.

Before closing the section, we would like to mention that
the GOE$^2$ distribution seems rather universal~\cite{IS2004ip} 
but the numerical values 
of statistical quantities of the GOE$^2$ have not been given 
in the literature. Since we have the data at hand, we list them here.
The known data for GOE/GUE/GSE\cite{TW1999} are also presented for 
comparison. (In this table, 
the data for GSE is shown according to the notation in
\cite{BR2001c}. Therefore the values of the average and standard 
deviation in GSE are different from those values  written in \cite{TW2000}.  
)

\begin{center}
 \begin{tabular}[t]{|l|c|c|c|c|}
  \hline
          & average  & s.d.   & skewness & kurtosis \\ \hline
  GOE$^2$ & -0.49364 & 1.1100 & 0.3917   & 0.309    \\ 
  GOE     & -1.20653 & 1.2680 & 0.2935   & 0.165    \\ 
  GUE     & -1.77109 & 0.9018 & 0.2241   & 0.093    \\ 
  GSE     & -3.26243 & 1.0176 & 0.1655   & 0.049    \\ \hline
 \end{tabular}
\end{center}

\setcounter{equation}{0}

\section{Transition around $F_0$}
\label{F0}
\subsection{Limiting Kernel}
Let us suppose that, when $\g_-=\g_0=1/\g_+$, 
the limit shape is tangent to that in bulk \eqref{adef} 
at $\b=\b_0$.  Let us call this point the $F_0$ point.
In this case we again have to consider the modified model
as in the proof of (i-c), (ii-a) and (iii-a) in Theorem 3.1
since the parameter of geometric distribution
at the ($i=1,j=1$) point is close to unity in the original model.
The relation between the modified and original  models is 
given in~\eqref{ModDef}, 
\begin{align}
&\quad\P[h(r_1,t=2N)<l_1,\cdots ,h(r_m,t=2N)<l_m]\notag\\ 
&=\P[h^+(r_1,2N)<l_1,\cdots ,h^+(r_m,2N)<l_m] + 
\frac{\g_+\g_-}{1-\g_+\g_-}
\big\{\P[h^+(r_1,2N)<l_1,\cdots ,h^+(r_m,2N)<l_m] \notag\\
&\quad-\P[h^+(r_1,2N)<l_1-1,\cdots ,h^+(r_m,2N)<l_m-1]
\big\}.
\label{ModDef2}
\end{align}
When we consider the case where $\g_+\g_-=1$ we always treat 
$\P[h^+(r_1,t)<l_1,\cdots ,h^+(r_m,t)<l_m]$ in the right hand side as 
the Fredholm determinant~\eqref{Fdet} with the 
kernel~\eqref{dif2}--~\eqref{psi}.
In the scaling limit where
\label{gpm-scale}
\begin{align}
&r_i=2\b_0 N+2\t_iN^{2/3},\notag\\ 
&l_i =a\left(\b_0+\frac{c(\b_0)\t_i}{N^{1/3}}\right)N
+d(\b_0)N^{\frac13}s_i, \notag\\ 
&\g_{\pm} = \frac{1}{\g_0}\left( 1-\frac{\o_{\pm}}{d(\b_0)N^{1/3}}\right),
\end{align}
one finds
\begin{align}
 & \lim_{N\to\i} \P[H_N(\t_1,\b_0) \leq s_1, \cdots , 
H_N(\t_m,\b_0) \leq s_m]\notag\\
&=
 \left(1+\frac{1}{\o_++\o_-}\sum_{j=1}^m \pd{s_j}\right)
\lim_{N\to\i} \P[H^+_N(\t_1,\b_0) \leq s_1, 
\cdots , H^+_N(\t_m,\b_0) \leq s_m]. 
\label{F0cor}
\end{align}

We obtain the kernel describing the 
multi-point height fluctuation near this $F_0$ point.
The result is
\begin{theorem}
\hspace{90pt}\\
When $\o_++\o_->0$,
\begin{equation}
\label{detKlim_F0}
 \lim_{N\to\i} \P[H_N(\t_1,\b_0) \leq s_1, \cdots , H_N(\t_m,\b_0) \leq s_m]
 =
 \left(1+\frac{1}{\o_++\o_-}\sum_{j=1}^m \pd{s_j}\right) \det(1+\K \G),
\end{equation}
where $\G(\t_j,\xi)=-\chi_{(s_j,\i)}(\xi)$ ($j=1,2,\cdots,m$), 
\begin{align}
 &\quad 
 \K(\t_1,\xi_1;\t_2,\xi_2)=
 \K_2(\t_1,\xi_1;\t_2,\xi_2) 
 +(\o_++\o_-)\mathcal{B}(\o_+,\t_1,\xi_1)\mathcal{B}'(w_-,\t_2,\xi_2),
\label{Th5.11}
\end{align}
and 
\begin{align}
&\mathcal{B}(\o_+,\t_1,\xi_1)=
\begin{cases}
  \int_0^{\i} d\l e^{-(\o_+-\t_1)\l}\Ai(\xi_1-\l),& 
  \o_+-\t_1>0, \\
   -\int_0^{\infty}d\l e^{(\o_+-\t_1)\l}\Ai(\xi_1+\l)
+e^{\frac{-\t_1^3+\o_+^3}{3}-\xi_1(\o_+-\t_1)},    & \o_+-\t_1<0, 
 \end{cases}
\\
&\mathcal{B}'(\o_-,\t_2,\xi_2)=
\begin{cases}
  \int_0^{\i} d\l e^{-(\o_-+\t_2)\l}\Ai(\xi_2-\l),& 
  \o_-+\t_2>0, \\
   -\int_0^{\i}d\l e^{(\o_-+\t_2)\l}\Ai(\xi_2+\l)
+e^{\frac{\t_2^3+\o_-^3}{3}-\xi_2(\o_-+\t_2)},    & \o_-+\t_2<0. 
 \end{cases}
\label{Th5.12}
\end{align}
\end{theorem}

\vspace{3mm}\noindent
{\bf Remarks.} 
\begin{enumerate}
\item 
The Fredholm representation~\eqref{detKlim_F0} 
and~\eqref{Th5.11} includes the terms which are not analytic 
for $\o_++\o_-<0$, i.e.,
the difficulty of divergence that we discussed in the proof of 
Theorem 3.1 (i-c), (ii-a) and (iii-a) arises. However, one can 
avoid this in a similar manner as for the Theorem 3.1.
  
For example let us consider the term included in 
this Fredholm determinant which corresponds to 
the last term in~\eqref{AnCo},
\begin{equation}
  (\o_++\o_-)e^{\o_+^3+\o_-^3}\int_s^{\i}d\xi e^{-\xi(\o_++\o_-)}.
\label{DaTe}
\end{equation}
For $\o_++\o_->0$, this can be calculated as
\begin{equation}
  (\o_++\o_-)e^{\o_+^3+\o_-^3}\int_s^{\i}d\xi e^{-\xi(\o_++\o_-)}
=e^{\o_+^3+\o_-^3-(\o_++\o_-)s}.
\label{RDaTe}
\end{equation}
Here though the LHS is meaningful only for $\o_++\o_->0$,
the RHS is analytic for all region of $\o_++\o_-$. 
Hence the difficulty of divergence of (\ref{DaTe}) can be avoided 
if one replaces the integral by the RHS in~\eqref{RDaTe} when
$\o_++\o_-<0$. Other terms which diverge when $\o_++\o_-<0$
can be treated in the same manner.

%
 
\item When we set $\o_{\pm}=0$ we obtain Theorem 3-1 (iii).
\end{enumerate}

\noindent
{\bf Proof.} 
For the case where $\o_+-\t_1>0, \o_-+\t_2>0$, the contour integral representation of the 
kernel reads
\begin{align}
 K_N(2u_1,x_1; 2u_2,x_2) 
 &=
 \frac{ (1-a)^{2(u_2-u_1)}}{(2\pi i)^2}
 \int_{C_{R_1}} \frac{dz_1}{z_1}\int_{C_{R_2}}\frac{dz_2}{z_2}
 \frac{z_2^{x_2}}{z_1^{x_1}} \frac{z_1}{z_1-z_2}
 \frac{(1-\a/z_1)^{N-1+u_1}(1-\a z_2)^{N-1-u_2}}
      {(1-\a z_1)^{N-1-u_1}(1-\a/z_2)^{N-1+u_2}}
 \notag\\
 &\quad \times
 \frac{1-\g_+ z_2}{1-\g_+ z_1} \frac{1-\g_-/z_1}{1-\g_-/z_2}.
\end{align}

The asymptotics can be studied in the same manner as in 
the previous section with the result,
\begin{align}
 K_N(2u_1,x_1; 2u_2, x_2) 
 &\sim
 (1-\a)^{2(u_2-u_1)} \big(p_c(\b_0)\big)^{(\xi_2-\xi_1)d(\b_0)N^{1/3}}
 \frac{1}{d(\b_0)N^{1/3}} \notag\\
 &\quad\times
 e^{ \l_1(\b_0)c(\b_0)N^{2/3}(\t_1-\t_2)
    +\l_2(\b_0)c(\b_0)^2 N^{1/3}(\t_1^2-\t_2^2)
    +\l_3(\b_0)c(\b_0)^3 (\t_1^3-\t_2^3)+\xi_2 \t_2-\xi_1 \t_1} \notag\\
 &\quad\times
 \frac{1}{4\pi^2}
 \int_{\text{Im}w_1=\eta_1} dw_1 \int_{\text{Im}w_2=\eta_2} dw_2 
e^{i\xi_1 w_1+i\xi_2 w_2+\frac{i}{3}( w_1^3+ w_2^3)}\notag\\
 &\quad\times\left(-\frac{1}{\tau_2-\tau_1+i(w_1+w_2)}
       +\frac{\o_++\o_-}{(\o_+-\t_1+iw_1)(\o_-+\tau_2+iw_2)}\right).
\label{F0K1}
\end{align}
The one arrives at the desired expression.

In the case of $\o_+-\t_1>0, \o_-+\t_2<0$ the contribution of the pole at
$z_2=\g_-$ are added as
\begin{align}
&\quad\frac{ (1-\a)^{2(u_2-u_1)}}{2\pi i}
 \int_{C_{R_1}} \frac{dz_1}{z_1} \frac{1}{z_1^{x_1}}
 \frac{(1-\a/z_1)^{N-1+u_1} \g_-^{x_2}}{(1-\a z_1)^{N-1-u_1}(1-\g_+z_1)}   
 \frac{(1-\a\g_-)^{N-1-u_2}(1-\g_+\g_-)}{(1-\a/\g_-)^{N-1+u_2}} \notag\\
&\sim
(1-\a)^{2(u_2-u_1)} (p_c(\b_0))^{(\xi_2-\xi_1)d(\b_0)N^{1/3}}
 \frac{1}{d(\b_0)N^{1/3}} \notag\\
 &\quad\times
 e^{ \l_1(\b_0)c(\b_0)N^{2/3}(\t_1-\t_2)
    +\l_2(\b_0)c(\b_0)^2 N^{1/3}(\t_1^2-\t_2^2)
    +\l_3(\b_0)c(\b_0)^3 (\t_1^3-\t_2^3)+\xi_2 \t_2-\xi_1 \t_1} \notag\\
 &\quad\times
 e^{\frac{\t_2^3+\o_-^3}{3}-\xi_2(\t_2+\o_-)}
\frac1{2\pi}\int_{\text{Im}w_1=\eta_1}
dw_1 
\frac{e^{i\xi_1 w_1+\frac{i}{3} w_1^3}}
{\o_+-\t_1+iw_1}
\label{F0K2}
\end{align}
~\eqref{F0K1} and~\eqref{F0K2} lead to the desired expression. Similar to 
this case we add the term of pole at $z_1=1/\g_+$ in the case of 
$\o_+-\t_1<0, \o_-+\t_2>0$. This term reads   
\begin{align}
&\quad\frac{ (1-\a)^{2(u_2-u_1)}}{2\pi i}
 \int_{C_{R_2}} \frac{dz_2}{z_2} z_2^{x_2}
 \frac{(1-\a z_2)^{N-1-u_2} \g_+^{x_1}}{(1-\a/z_2)^{N-1+u_2}(\g_+z_2-1)}   
 \frac{(1-\a\g_+)^{N-1+u_1}(1-\g_+\g_-)}{(1-\a/\g_+)^{N-1-u_1}}\notag\\
&\sim
(1-\a)^{2(u_2-u_1)} (p_c(\b_0))^{(\xi_2-\xi_1)d(\b_0)N^{1/3}}
 \frac{1}{d(\b_0)N^{1/3}} \notag\\
 &\quad\times
 e^{ \l_1(\b_0)c(\b_0)N^{2/3}(\t_1-\t_2)
    +\l_2(\b_0)c(\b_0)^2 N^{1/3}(\t_1^2-\t_2^2)
    +\l_3(\b_0)c(\b_0)^3 (\t_1^3-\t_2^3)+\xi_2 \t_2-\xi_1 \t_1} \notag\\
 &\quad\times e^{\frac{-\t_1^3+\o_+^3}{3}-\xi_1(\o_+-\t_1)}
\frac1{2\pi}\int_{\text{Im}w_2=\eta_2}
dw_2 
\frac{e^{i\xi_1 w_2+\frac{i}{3} w_2^3}}
{\o_-+\tau_2+iw_2}
\label{F0K3}
\end{align}

Finally we consider the case of $\o_+-\t_1<0, \o_-+\t_2<0$. In addition to
~\eqref{F0K1},~\eqref{F0K2} and \eqref{F0K3}  we have to consider the term
 caused by the product of the residues,
\begin{align}
&\quad (1-\a)^{2(u_2-u_1)}\g_+^{x_1}\g_-^{x_2}\frac{(1-\a\g_+)^{N-1+u_1}}{(1-\a/\g_+)^{N-1-u_1}}
\frac{(1-\a\g_-)^{N-1-u_2}}{(1-\a/\g_-)^{N-1+u_2}}
\notag\\
&\sim
(1-\a)^{2(u_2-u_1)} (p_c(\b_0))^{(\xi_2-\xi_1)d(\b_0)N^{1/3}}
 \frac{1}{d(\b_0)N^{1/3}} \notag\\
&\quad\times
 e^{ \l_1(\b_0)c(\b_0)N^{2/3}(\t_1-\t_2)
    +\l_2(\b_0)c(\b_0)^2 N^{1/3}(\t_1^2-\t_2^2)
    +\l_3(\b_0)c(\b_0)^3 (\t_1^3-\t_2^3)+\xi_2 \t_2-\xi_1 \t_1} \notag\\
&\quad\times
  e^{\frac{\t_2^3+\o_-^3}{3}-\xi_2(\t_2+\o_-)}
  e^{\frac{-\t_1^3+\o_+^3}{3}-\xi_2(\o_+-\t_1)}.
\end{align}
Combining each term we get the desired expression.
\qed



\subsection{Connection to the Baik-Rains analysis}
When we specialize to the case of the one-point height fluctuation, 
our formula again should be equivalent to the ones obtained in 
\cite{BR2000}. 
We show the equivalence in this subsection.

When we set $\t_i=0$, the Fredholm determinant reads 
\begin{equation}
  \det(1+\K\G) 
=
 F_2(s)
\left\{1-(\o_++\o_-)\int_s^{\i}dx 
\int_s^{\i}dy \r(x,y)B(x,\o_+) B(y,\o_-)\right\},
\end{equation}
where $F_2(s)$ is the GUE Tracy-Widom distribution and $B(x,\o)$ is defined 
in~\eqref{defB}. We have the following proposition for this expression.
\begin{proposition}
\begin{equation}
1-(\o_++\o_-)\int_s^{\i}dx \int_s^{\i}dy \r(x,y)B(x,\o_+) B(y,\o_-)
=
a(s,\o_+)a(s,\o_-)-b(s,\o_+)b(s,\o_-) .
\end{equation}
\end{proposition}
\vspace{3mm}\noindent
{\bf Remark.} 
Using the proposition and~\eqref{F0cor}, 
we can easily calculate the one-point height fluctuation. 
\begin{align}
   & \lim_{N\to\i} \P[H_N(\t=0,\b_0) \leq s]\notag\\
&=
 \left(1+\frac{1}{\o_++\o_-}\pd{s}\right)
\big\{a(s,\o_+)a(s,\o_-)-b(s,\o_+)b(s,\o_-)\big\}F_2(s)\notag\\
&=
a(s,\o_+)a(s,\o_-)F_2(s)+\frac1{\o_++\o_-}
\big\{a(s,\o_+)a(s,\o_-)-b(s,\o_+)b(s,\o_-)\big\}F_2'(s).  
\end{align}
This corresponds to the expression of the one-point fluctuation
in~\cite{BR2000}.

\medskip
\noindent  
{\bf Proof.} 
Let $f$ denote the function on the left hand side.
By differentiation, one finds
\begin{equation}
 \pd{s} f(s)=(\o_++\o_-)\big\{a(s,\o_+)a(s,\o_-)-f(s)\big\}.
\end{equation}
This can be read as
\begin{equation}
 f(s) 
 =
 a(s,\o_+)a(s,\o_-)+\frac{\partial}{\partial s}
 \int_s^{\i}dx \int_s^{\i}dy \r(x,y)B(x,\o_+) B(y,\o_-).
\end{equation}
The second term is now computed as
\begin{align}
&\frac{\partial}{\partial s}
\int_s^{\i}dx \int_s^{\i}dy \r(x,y)B(x,\o_+) B(y,\o_-)\notag\\
&=-B(s,\o_+)\int_s^{\i}dy \r(s,y)B(y,\o_-)+ 
\int_s^{\i}dx B(x,\o_+)\frac{\partial}{\partial s}
\int_s^{\i}dy \r(x,y)B(y,\o_-) \notag\\
&=-B(s,\o_+)\int_s^{\i}dy \r(s,y)B(y,\o_-)\notag\\
&\quad+ 
\int_s^{\i}dx B(x,\o_+)
\int_s^{\i}dy \big\{-\r(x,s)+\delta_+(x-s)\big\}\r(y,s)B(y,\o_-)\notag\\
&=-\int_s^{\i}dx \r(x,s)B(x,\o_+)\int_s^{\i}dy \r(y,s)B(y,\o_-) \notag\\
&=-b(s,\o_+)b(s,\o_-).
\end{align}
where we use the equation (2.16) in~\cite{TW1994}.
\qed


\section{Conclusion}
We have studied the PNG model with external sources. 
We have represented the multi-point equal time correlation functions 
of the height fluctuation as the Fredholm determinant.
In order to get these quantities we have first identified the
limit shape of the model. The results are summarized in Theorem 2.1. 
The limit shape consists of a straight line and a circular one.
There are three cases according to the value of $\g_+\g_-$ and 
they are illustrated in Fig.2.  
The statistics of the one-point height fluctuation is 
Gaussian on the straight line and the GUE Tracy-Widom distribution
on the circular line.
In addition, there are special points denoted by $\b_{\pm}$ where 
the two limit shapes meet and their statistics of the height fluctuation
is GOE$^2$ when $\g_+\g_-<1$ or $F_0$ when $\g_+\g_-=1$. 


Our main results are summarized as Theorems 3.1, 4.1 and 5.1.
The spatial configuration of the height 
fluctuation over some region of the surface has been shown to 
converge to the stochastic processes represented by the 
Fredholm determinants obtained in the above theorems.
In addition to the one-dimensional Brownian motion and the Airy process
which describe the correlations near edges and in the bulk respectively, 
we have obtained the Fredholm determinant which describes the 
correlation near the points $\b_\pm$. 

These quantities  vary with the parameter of the nucleation at edges $\g_{\pm}$
in \eqref{aPNG} and~\eqref{bPNG}. 
For the case of  $\g_+\g_-<1$ the Fredholm determinant obtained in Theorem 4.1
describes the transition between GOE$^2$ and GUE/Gaussian.   
On the other hand, when
$\g_+\g_-=1$ the points $\b_\pm$ merge to $\b_0$ and 
we have obtained the Fredholm determinant expression with the transition kernel between $F_0$ and Gaussian.
These Fredholm determinants have the parameters $\t_i$ which determine 
the position in the surface. 
When we set $\t_i=\t\neq 0$  we can obtain the one-point 
distribution at the point away from $\b_{\pm}$.  
Fig.4 shows the comparison of 
these transitive distributions
with the Monte-Carlo simulations of the PNG model in the corresponding points.
These figures confirm our analyses.

\section*{Acknowledgment}
The authors would like to thank M. Katori and T. Nagao
for fruitful discussions and comments. 
They also thank the Yukawa Institute for Theoretical Physics at Kyoto
University,  
where this study was influenced by the stimulating discussions 
during the YITP-W-03-18 on `` Stochastic models in statistical mechanics''. 
The work of T.I. is partly supported by the Grant-in-Aid for JSPS
Fellows, the Ministry of Education, Culture, Sports, Science and 
Technology, Japan.
The work of T.S. is partly supported by the Grant-in-Aid for Young 
Scientists (B), the Ministry of Education, Culture, Sports, Science and 
Technology, Japan.


\newpage
\begin{large}
\noindent
Figure Captions
\end{large}

\vspace{10mm}
\noindent
Fig.1:
Rules of the discrete PNG model. Fig. (a) shows 
an example of the rules 1 and 2.
A step with height $k$ at the origin is generated 
according to the geometric distribution.
It grows laterally one step in both directions in the next time step.

Fig. (b) shows a collision of the steps with height one and two. The height
of the origin, which is the point they collide, is two by the rule 3. 


\vspace{10mm}
\noindent
Fig.2:
Typical examples of the limit shape of the PNG model with external sources.
The solid lines show the snapshots of Monte Carlo simulations for
$N=5000, \a=0.32$. The dashed 
lines represent the limit shape in bulk given by~\eqref{adef}.
The parameters $\g_+$ and $\g_-$ are $\g_+=0.79 ,\g_-=0.63$ in (a),
$\g_+=1.58 ,\g_-=1.26$ in (b) and $\g_+=1.58 ,\g_-=0.63$ in (c).

Fig.(a) corresponds to the region (i) in Theorem 2.1 where the 
hight fluctuation in bulk obeys the GUE Tracy-Widom distribution
while the fluctuations near both edges are described by Gaussian.
Fig.(b) corresponds to the region (ii) 
where the height fluctuations at all points obey the Gaussian distribution.
Fig.(c) represent the case of region (iii).
The limit shape is a straight line  tangent to the limit 
shape in bulk. 
 
\vspace{10mm}
\noindent
Fig. 3: 
Average in the GOE$^2$ to GUE/Gaussian transition.
Three lines represent the asymptotics in the case where 
$\t\sim \infty, 0, -\infty$  
respectively. $+$ gives the value obtained by 
solving \eqref{a_s}--\eqref{b_o} numerically.

\vspace{10mm}
\noindent
Fig4: The limiting distribution near the competing points between bulk and edge. 
In Fig.(a) the three curves are the distributions for $\tau=1,0,-1$ from the left
obtained in Theorem 4.1. Each point is the data of a Monte-Carlo simulation near the origin where 
$t=2000(N=1000), \a=0.32, \g_+=0.32, \g_-=1.0 $ and $x=352,0,-352$ respectively.
Fig.(b) shows the distribution near $F_0$. The curves  represent the 
transitive distributions for $\t=0,1$ from the left using Theorem5.1. 
The points are numerical data for  
$t=2000(N=1000), \a=0.32, \g_+=1.0 , \g_-=1.0 $ and $x=0,352$ respectively.
\newpage
\renewcommand{\thepage}{Figure 1}
\begin{picture}(400,700)
\put(50,700){(a)}
\put(130,645){rule 1}
\put(225,645){rule 2}
\put(50,600){\resizebox{10cm}{2cm}{\includegraphics{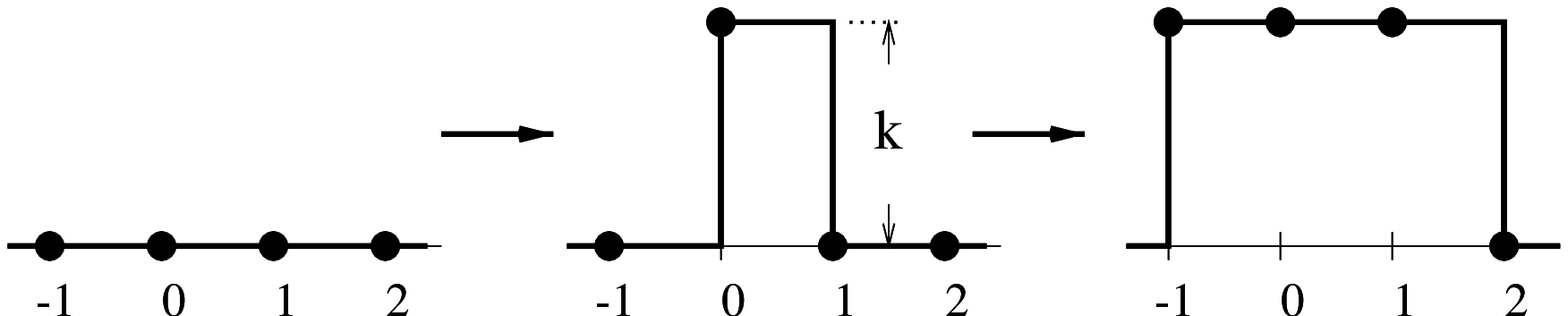}}}
\put(50,560){(b)}
\put(180,480){rule 3}
\put(50,460){\resizebox{10cm}{2cm}{\includegraphics{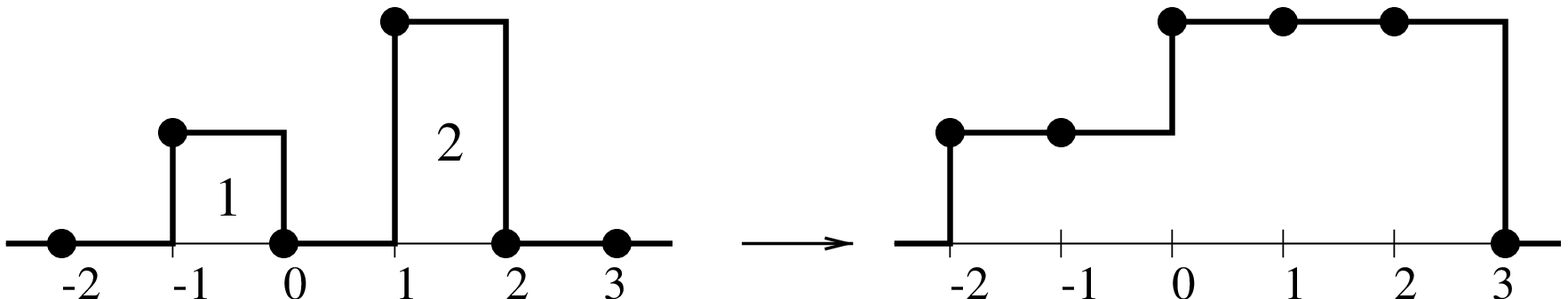}}}
\end{picture}

\newpage
\renewcommand{\thepage}{Figure 2}
\begin{picture}(400,700)
\put(30,680){(a)}
\put(320,490){$r$}
\put(40,660){$h$}
\put(50,500){\resizebox{10cm}{6cm}{\includegraphics{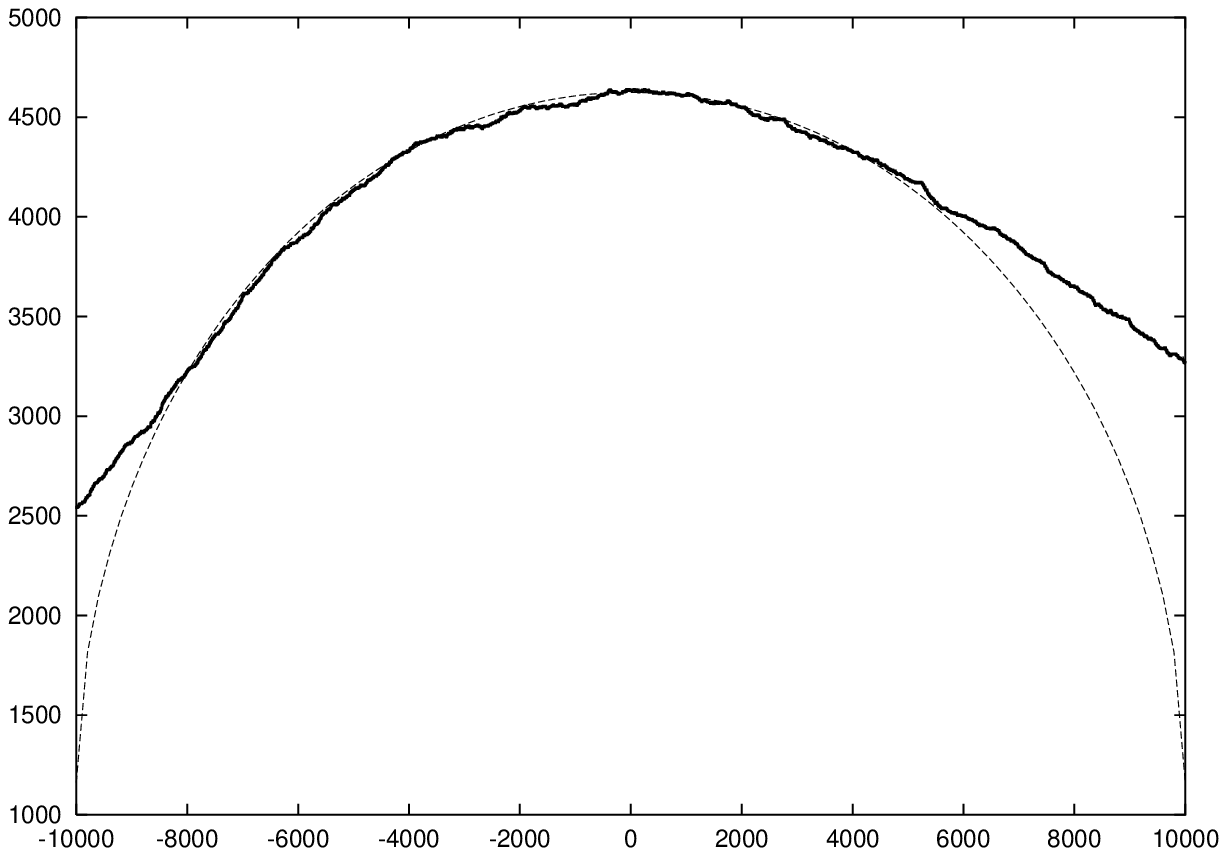}}}
\put(30,480){(b)}
\put(320,290){$r$}
\put(40,460){$h$}
\put(50,300){\resizebox{10cm}{6cm}{\includegraphics{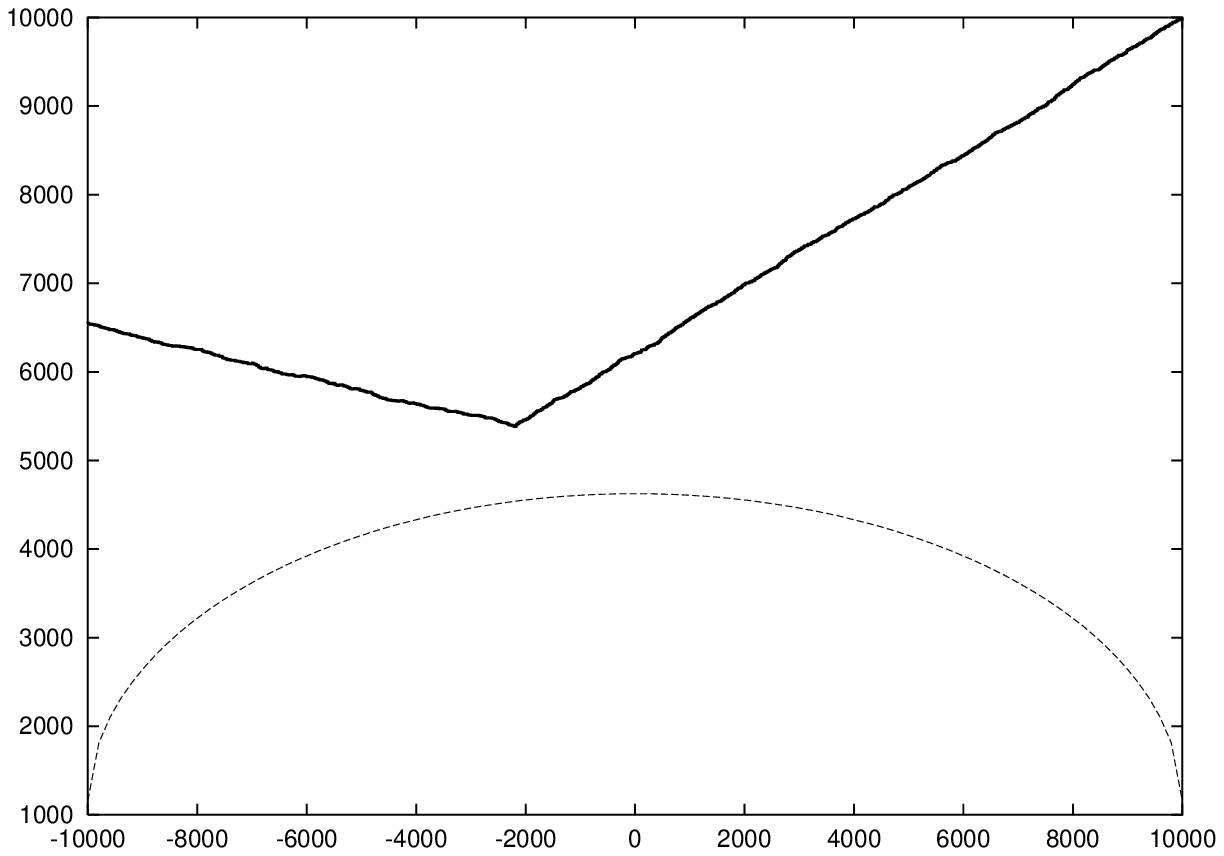}}}
\put(30,280){(c)}
\put(320,90){$r$}
\put(40,260){$h$}
\put(50,100){\resizebox{10cm}{6cm}{\includegraphics{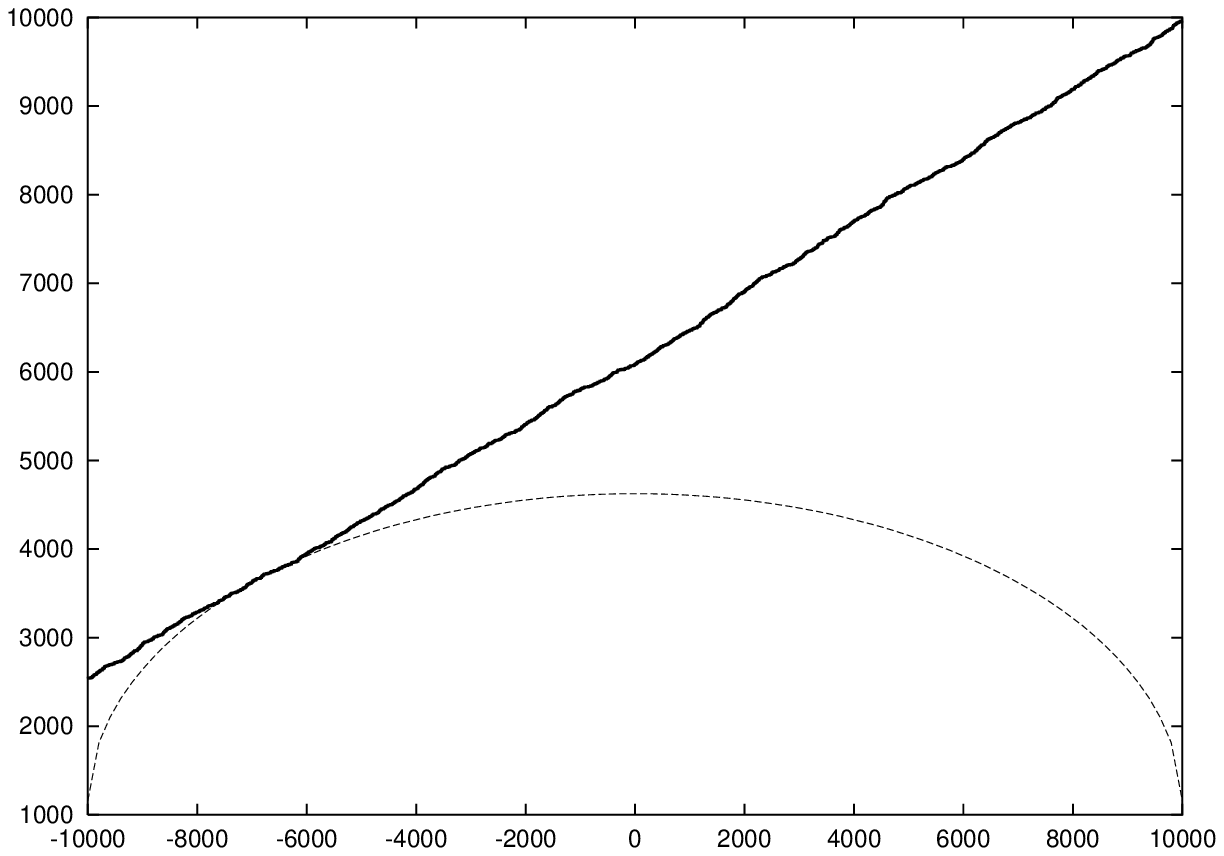}}}
\end{picture}
\newpage
\renewcommand{\thepage}{Figure 3}
\begin{picture}(400,300)
\put(50,0){\includegraphics{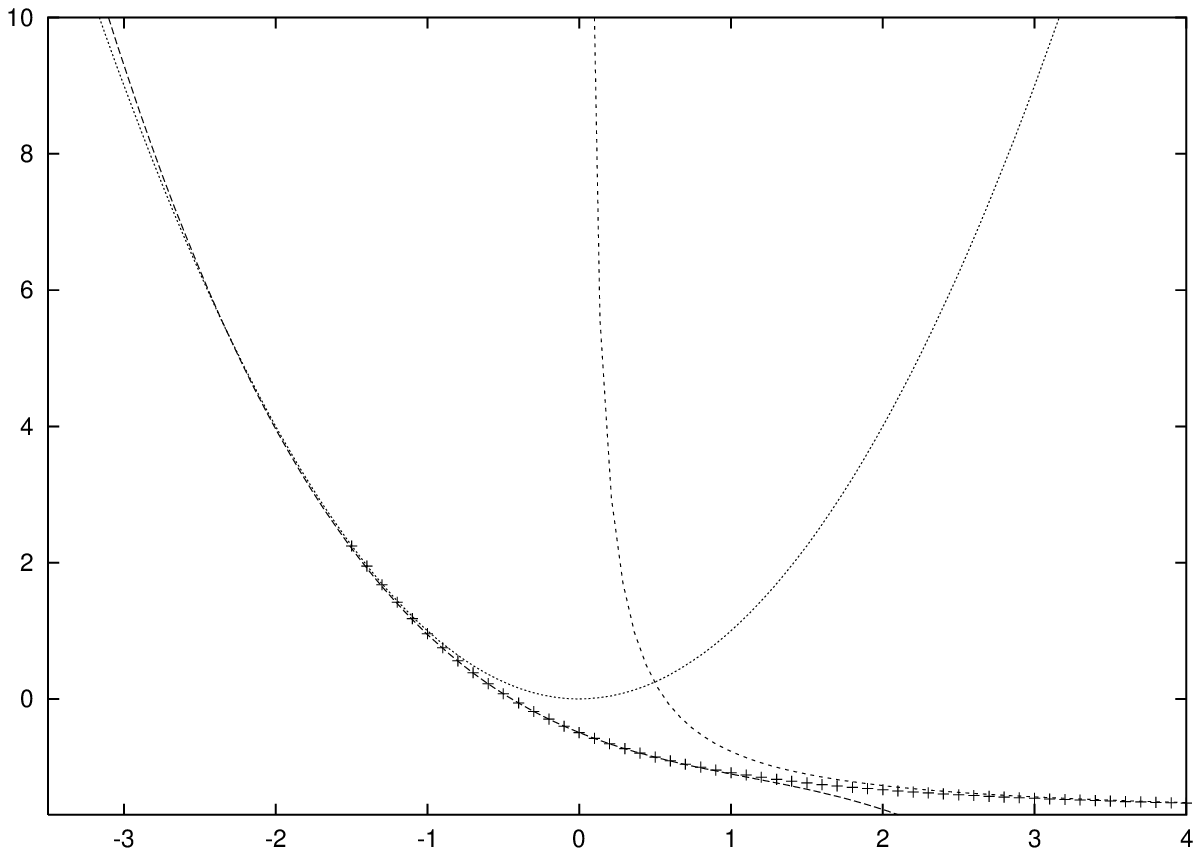}}
\put(230,-5){$\t$}
\put(230,230){$\t\to\i$}
\put(350,200){$\t\to-\i$}
\put(95,230){$\t\sim 0$}
\put(70,250){average}
\end{picture}
\newpage
\renewcommand{\thepage}{Figure 4}
\begin{picture}(400,700)
\put(50,680){(a)}
\put(50,500){\resizebox{10cm}{6cm}{\includegraphics{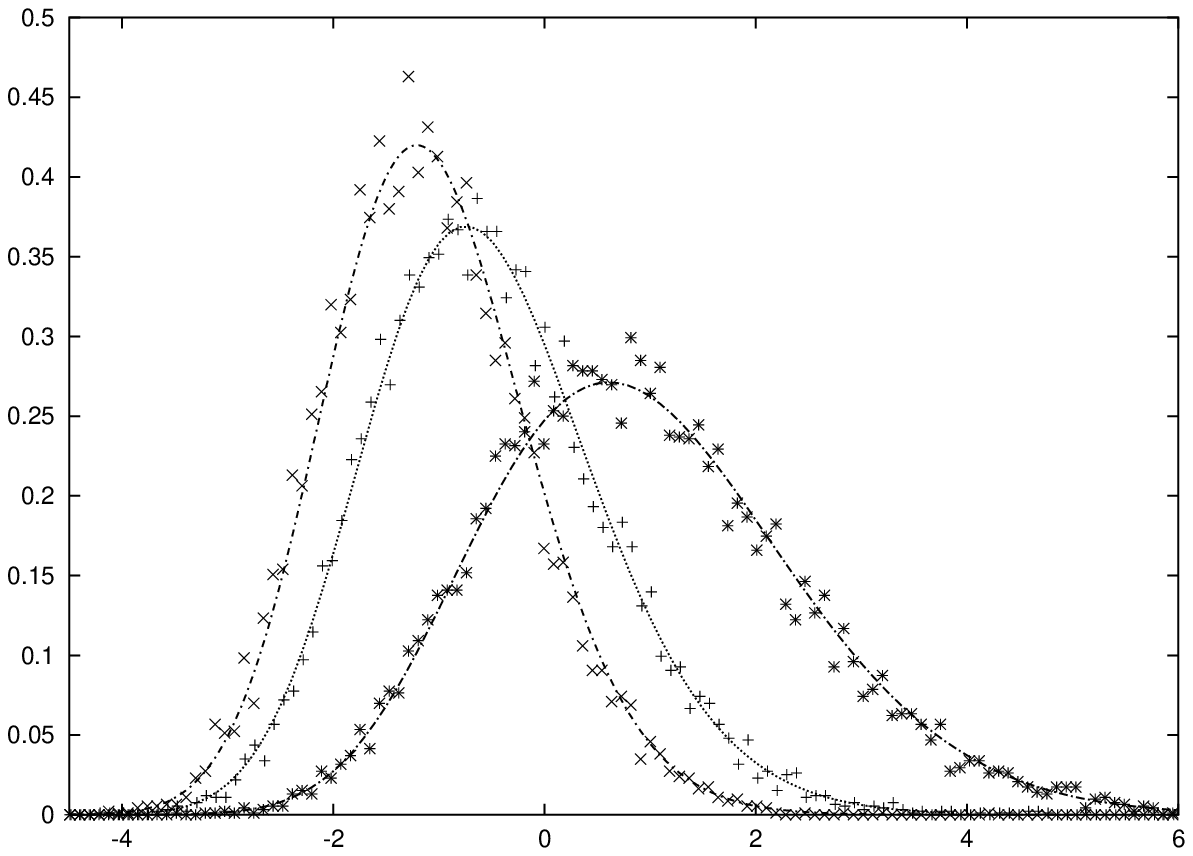}}}
\put(190,490){$s$}
\put(100,630){$\t=1$}
\put(170,630){$\t=0$}
\put(180,620){\small{GOE$^2$}}
\put(240,570){$\t=-1$}
\put(50,480){(b)}
\put(50,300){\resizebox{10cm}{6cm}{\includegraphics{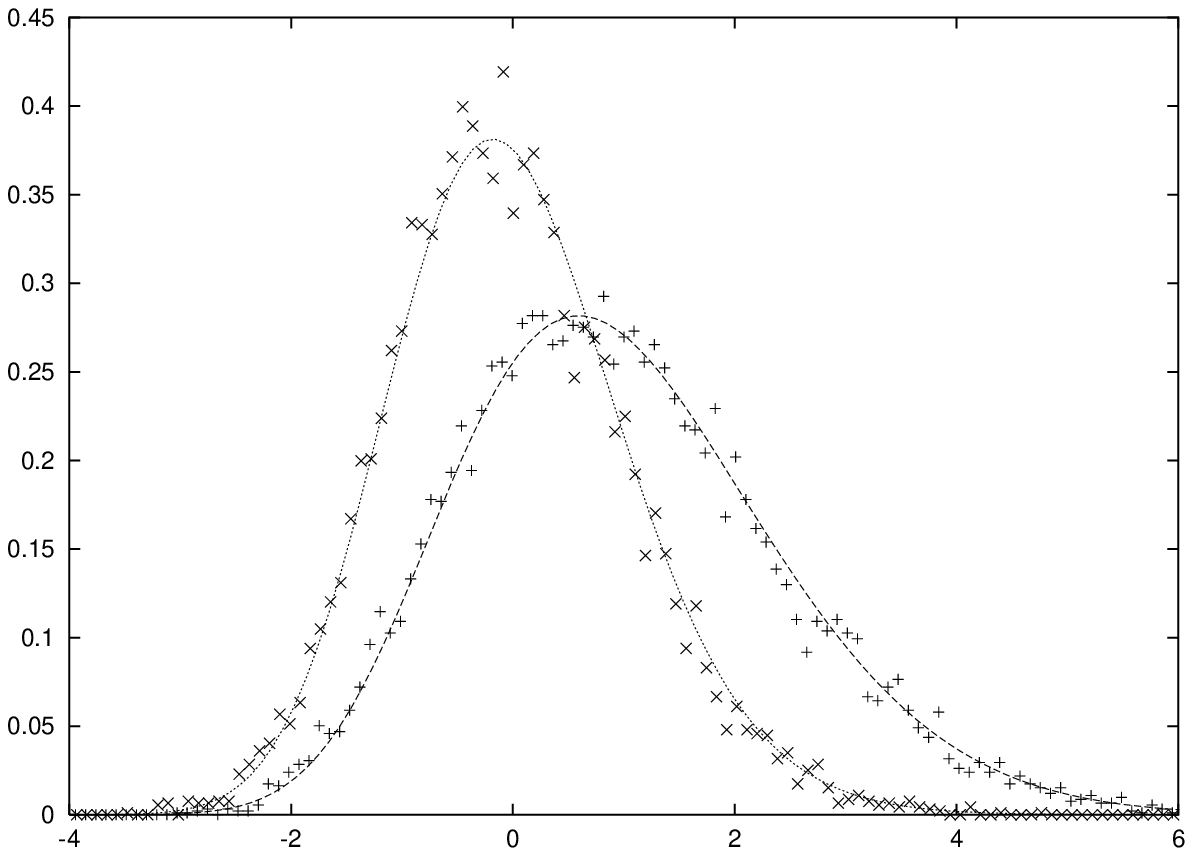}}}
\put(190,290){$s$}
\put(120,430){$\t=0$}
\put(125,420){\small{$F_0$}}
\put(240,370){$\t=1$}
\end{picture}
\end{document}